\documentclass[12pt]{article}

\newlength{\defbaselineskip}
\usepackage[colorlinks,linkcolor=blue,anchorcolor=blue,citecolor=blue]{hyperref}
\usepackage{float,caption,graphicx,amsmath,bbold,multirow,booktabs,threeparttable,multirow,footmisc,amsthm,diagbox,longtable,mathrsfs,upgreek,adjustbox,cite,subfig,rotating,colortbl,bm,xr}
\usepackage[table]{xcolor}
\usepackage[title]{appendix}
\usepackage[authoryear,comma,sort&compress]{natbib}
\theoremstyle{break}
\newtheorem{rem}{\textbf{Remark}}
\newtheorem{thm}{\textbf{Theorem}}[section]

\newtheorem{cor}{\textbf{Corollary}}[section]

\newtheorem{ass}{\textbf{Assumption}}[section]
\newtheorem{cond}{\textbf{Condition}}

\numberwithin{equation}{section}

\usepackage[graphicx]{realboxes}
\allowdisplaybreaks[4]

\begin{document}
\baselineskip=0.8 true cm

\title{Inference for the panel ARMA--GARCH model when both $N$ and $T$ are large\footnote{The codes for the numerical analysis are accessible at \href{https://github.com/subingGitHub/Codes-for-panel-ARMA-GARCH-model/}{https://github.com/subingGitHub/Codes-for-panel-ARMA-GARCH-model}.}}
\author{Bing Su and Ke Zhu\footnote{Address correspondence to Ke Zhu: Department of Statistics \& Actuarial Science, University of Hong Kong, Hong Kong; E-mail:  mazhuke@hku.hk} \\\\
\emph{University of Hong Kong}}
\date{}
\maketitle

%Estimation of panel ARMA model with fixed effects and GARCH disturbances

%%%%%%%%%%%%%%%%%%%%%%%%%%%%%%%%%%%%%%%%%%%%%%%%%%%%%%%%%%%
\begin{center}
\begin{minipage}{15truecm}	
{\bf Abstract. }
We propose a panel ARMA--GARCH model to capture the dynamics of large panel data with $N$ individuals over $T$ time periods.
For this model, we provide a two-step estimation procedure to estimate the ARMA parameters and GARCH parameters stepwisely.
Under some regular conditions, we show that all of the proposed estimators are asymptotically normal with
the convergence rate $(NT)^{-1/2}$, and they have the asymptotic biases when both $N$ and $T$ diverge to infinity at the same rate.
Particularly, we find that the asymptotic biases result from the fixed effect, estimation effect, and unobservable initial values.
To correct the biases, we further propose the bias-corrected version of estimators by using either the analytical asymptotics or jackknife method.
Our asymptotic results are based on a new central limit theorem for the linear-quadratic form in the martingale difference sequence, when the weight matrix is uniformly bounded in row and column.
Simulations and one real example are given to demonstrate the usefulness of our panel ARMA--GARCH model.\\

\end{minipage}
\end{center}

\newpage

%%%%%%%%%%%%%%%%%%%%%%%%%%%%%%%%%%%%%%%%%%%%%%%%%%%%%%%%%%%
\section{Introduction}\label{section_introducation}

The dynamic panel model is a valuable statistical tool for analyzing panel data and investigating the dynamic relationships between variables, making it useful in various fields such as economics, finance, and social sciences.
See \cite{Baltagi2021} and \cite{Hsiao2022} for its surveys.
%its empirical elaborations in \cite{Baltagi1986}, \cite{Millimet2017}, and \cite{Tsionas2019}, and
Let $(y_{it},x_{it})\in\mathcal{R}\times \mathcal{R}^{D_x}$ be a pair of observations for individual unit $i$ at time point $t$, where $i=1,...,N$ and $t=1,...,T$. Here,
$N$ is the number of cross-sectional units and $T$ is the number of time periods. To
describe a first-order autoregressive (AR(1)) structure of $y_{it}$ as well as a
linear relationship between $y_{it}$ and the $D_x$-dimensional vector of exogenous variables $x_{it}$, the benchmark dynamic panel model is defined as
\begin{equation}\label{model_dynamic_panel}
y_{it} = \mu_{0i} + x_{it}'\beta_0 +  \phi_0 y_{i,t-1} + u_{it},
\end{equation}
for $i=1,...,N$ and $t=1,...,T$, where $\mu_{0i}$ is the unobservable fixed effect to characterize the individual heterogeneity, $\beta_0$ is a
$D_x$-dimensional vector of unknown parameters, $\phi_0$ is an unknown scalar parameter, and $u_{it}$ is the model disturbance.
Notably, $\mu_{0i}$ not only helps to mitigate issues related to omitted variable bias and endogeneity, but also
holds practical significance. For example, if $y_{it}$ is the return of asset $i$ at time $t$, $\mu_{0i}$ represents the excess return of this asset over $D_x$ different risk factors $x_{it}$, and investors have a preference for larger values of $\mu_{0i}$. In this context, the statistical inference of $\mu_{0i}$ is practically useful.

Although the presence of $\mu_{0i}$ is crucial, it causes the well-known incidental parameter problem (\citealp{NS1948}), making the estimation of model (\ref{model_dynamic_panel}) challenging. To solve this problem, researchers have explored two strands of literature.
The first strand focuses on the least square (LS) estimator (also known as within estimator) of $\beta_0$ and $\phi_0$ by concentrating out $\mu_{0i}$. When $T$ is fixed, \cite{Nickell1981} reveals that the LS estimator is biased. When both $N$ and $T$ diverge to infinity at the same rate, the asymptotic bias of LS estimator appears and it can be corrected by the analytical asymptotics method in \cite{Hahn2002} or the Jackknife method in \cite{HN2004} and \cite{Dhaene2015}.
%; see also other bias correction methods in \cite{GPY2010} and \cite{GK2015}.
Unlike the LS estimation method, the estimation method in the second strand removes $\mu_{0i}$ by adopting the first difference (FD) treatment. This leads to the FD-based generalized method of moments (FD-GMM) estimator in \cite{Arellano1991}  and the FD-based maximum likelihood (FD-ML) estimator in  \cite{Hsiao2002}. Despite FD-GMM and FD-ML estimators of $\beta_0$ and $\phi_0$ being consistent for fixed $T$ and large $N$, they are highly model specific and not applicable for estimating $\mu_{0i}$.
In addition, \cite{AA2003} illustrates that similar to the LS estimator, the FD-GMM estimator also suffers an asymptotic bias when both $N$ and $T$ diverge to infinity at the same rate.

%and \cite{Ahn1995}

Needless to say, model (\ref{model_dynamic_panel}) is inadequate to capture the higher-order serial correlation as well as conditional heteroskedasticity of $y_{it}$. To remedy this deficiency, we propose a panel autoregressive moving-average (ARMA) model
with fixed effects and generalized autoregressive conditional heteroskedasticity (GARCH) disturbances to study $y_{it}$,
where all of $y_{it}$ share the same ARMA and GARCH parameters cross-sectionally, but remain the unobservable fixed effects in both
panel ARMA and panel GARCH specifications. In short, our proposed model is termed as the panel ARMA--GARCH model. Clearly, the panel ARMA specification is applied to characterize the higher-order serial correlation of $y_{it}$, and the existence of MA part
could avoid the use of a large AR specification. The panel GARCH specification inspired by \cite{Engle1982} and \cite{Bollerslev1986} is introduced to depict the conditional heteroskedasticity, which is a prevalent phenomenon in financial and economic data. Although the ARMA--GARCH is a benchmark specification for studying the time series data (\citealp{FZ2004}; \citealp{ZL2011}), it has not been well explored in the dynamic panel framework.
When $y_{it}$ has the higher-order AR structure, a few works in
\cite{Hansen2007}, \cite{Lee2012}, and \cite{Lee2018} study the estimation of dynamic panel model. However, their estimation methods neither incorporate the MA specification for $y_{it}$ nor account for the GARCH disturbances. Till now, the only formal attempt for studying the panel GARCH specification is made by \cite{Pakel2011}. However, the estimation method in \cite{Pakel2011} has two major limitations. First, it only works for the small $N$ case with $N/T\to 0$, since it does not investigate the asymptotic bias of estimator when both $N$ and $T$ diverge to infinity at the same rate. Second, it assumes the zero-mean of $y_{it}$, so how the estimation of fixed effects and ARMA parameters affects that of GARCH specification is unclear.

%%% network garch ??

This paper is motivated to comprehensively study the estimation of the panel ARMA--GARCH model.
Our contributions to the literature are summarized as follows:

First, we propose a two-step estimation for the panel ARMA--GARCH model. To be specific, we estimate the panel ARMA specification by using the LS estimation method at step one. Based on the residuals from the step one, we then estimate the panel GARCH specification by adopting the variance-targeting quasi-maximum likelihood (VT-QML) estimation method (\citealp{Francq2011}) at step two. Under some regularity conditions, we prove that both LS estimator of ARMA parameters and VT-QML estimator of GARCH parameters are asymptotically normal with
the convergence rate $(NT)^{-1/2}$, and they have the asymptotic biases when both $N$ and $T$ diverge to infinity at the same rate.
Particularly, we illustrate that the existence of fixed effects and unobservable initial values produces the asymptotic biases in both LS and VT-QML estimators, and the latter estimator also suffers from the asymptotic bias caused by the first-step estimation effect. Moreover, we apply either the analytical asymptotics or jackknife method to correct the bias, and establish the related asymptotics for the bias-corrected estimators. For the fixed effects in both ARMA and GARCH specifications, we show that their estimators are asymptotically normal with the convergence rate $T^{-1/2}$, so their statistical inference can be implemented in a straightforward manner.

Second, we provide a new tool to study the convergence rate and central limit theorem (CLT) for the linear-quadratic form $V'MV+b'V$, where
$V=[v_{s}]$ is a vector of random variables, and $M=[m_{ss^*}]$ and $b=[b_{s}]$ are non-stochastic weight matrix and vector, respectively.
When $\{v_{s}\}$ is an independent and identically distributed (i.i.d.) or martingale difference sequence, \cite{Whittle1964}, \cite{Giraitis1997,Giraitis1998}, and \cite{Hsing2004} establish the CLT for the quadratic form (i.e., $b=0$) under the assumption that $m_{ss^*}$ is a function of $|s-s^*|$, while
this assumption on $M$ is further relaxed by \cite{Wu2007} and \cite{Giraitis2017}. When $\{v_{s}\}$ is an i.i.d. sequence, \cite{Kelejian2001} provides the CLT for the linear-quadratic form under a weaker assumption that $M$ is uniformly bounded in row and column. However, all of the aforementioned results of CLT are invalid for deriving the asymptotics of our proposed estimators. In this paper, under
the assumption that $M$ is uniformly bounded in row and column, we derive the CLT for the linear-quadratic form under general conditions on $V$, which allow $\{v_{s}\}$ to be block-independent with certain temporal dependence (that is, the martingale difference structure) within the block. Our new results on the linear-quadratic form are interesting in their own rights and could be useful for many other studies.

Third, our panel ARMA--GARCH model paves a new way to study the high-dimensional time series. In the literature,
the majority of work focuses on the estimation of high-dimensional vector AR model under certain regularity constraints; see, for example, the sparsity constraint in \cite{Basu2015}, \cite{KC2015}, and \cite{WuWu2016}, the banded constraint in \cite{Guo2016}, the network constraint in \cite{Zhu2017}, and the low rank constraint in \cite{BLM2019} and \cite{Wang2022}. See also
\cite{WBBM2023} for the exploration of  high-dimensional vector ARMA model under sparsity constraint.
%However, except for \cite{WuWu2016}, the aforementioned studies need the condition of i.i.d. disturbance for deriving the estimation consistency.
Our panel ARMA specification essentially poses a panel constraint that all ARMA parameters are the same cross-sectionally.
%where its disturbance is not necessarily i.i.d. and allowed to be an MDS or mixing sequence.
This panel constraint not only allows us to pool all information available in the panel, but also enables us to derive the asymptotic normality of LS estimator. Note that when $N$ diverges to infinity, so far only \cite{Zhu2017} establishes the asymptotic normality of model estimator, however, it neither accounts for the fixed effects nor considers the MA structure and GARCH disturbances. Compared with all of above studies on the high-dimensional time series models, our asymptotic normality result allows for more general martingale difference disturbances in the panel ARMA specification, thereby expanding the potential scope of applications for practitioners. Analogously, our panel GARCH specification also has a panel constraint that all GARCH parameters are the same cross-sectionally, as made by \cite{Pakel2011}. When $N$ diverges to infinity,
so far there is no theoretical development for the high-dimensional GARCH model.
%although some of multivariate GARCH models with a fixed $N$ have been empirically used (\citealp{Engle2019}; \citealp{Pakel2021}).
Our asymptotic normality result of the VT-QML estimator fills this gap for the first time in the literature. Owing to the panel structure in GARCH specification, the asymptotic normality of the VT-QML estimator works
for the case of moderate $T$ (say, e.g., $T=100$), as demonstrated by our simulation studies.
This overcomes a shortcoming of the classical GARCH models, which typically require a very large $T$ to get accurate estimates.
From a practical viewpoint, how to deal with the moderate $T$ case is important for studying many low-frequency real data.

%\cite{Kiefer1980} and \cite{Bai2021} consider the GLS estimation for the panel model without specifying the covariance structure. The GLS estimator has the closed-form expression containing exact moments of the disturbance, then its asymptotics can be established easily. It is expected to be less efficient than the ML estimator, and the generalization into the dynamic panel model with specified disturbances is challenging.

The remaining paper is organized as follows. Section \ref{section_model} introduces the panel ARMA-GARCH model and its two-step estimation method.  Section \ref{section_LQ} provides some general theoretical results for the linear-quadratic form. Section \ref{section_asymptotics} establishes the asymptotics of all proposed estimators and their corrected versions. Simulations are given in Section \ref{section_simulation}.  A real application is presented in Section \ref{section_example}. Concluding remarks are offered in Section \ref{section_conclusion}.
Technical proofs and some additional simulations are deferred into the supplementary materials.

Throughout the paper,  $\mathcal{R}$ is the one-dimensional Euclidean space, $I_n$ is the $n\times n$ identity matrix, and $l_n$ is the $n$-dimensional vector of ones.
For a matrix $A=[a_{ij}]\in\mathcal{R}^{p\times q}$,
$A'$ is its transpose, ${\rm tr}(A)$ is its trace,
$\|A\|_{\infty}$ is its $L_{\infty}$-norm, and $A^{-1}$ is its inverse when $p = q$.
For a random variable $\xi\in\mathcal{R}$, $||\xi||_p=({\rm E}|\xi|^p)^{1/p}$ is its $L_p$-norm for $0<p<\infty$. A sequence of matrices $\{A_{i}\}$ is uniformly bounded in row (or column) if $\sup_{i}||A_{i}||_{\infty}<\infty$ (or $\sup_{i}||A_{i}'||_{\infty}<\infty$).
A sequence of random variables $\{\xi_{i}\}$ is uniformly $L_p$-bounded if $\sup_{i} ||\xi_{i}||_p<\infty$.
Moreover, $O(1)$ denotes a generic bounded constant, $o_p(1) (O_p(1))$ denotes a sequence of random variables converging to zero (bounded) in probability, ``$\stackrel{p}{\longrightarrow}$'' denotes convergence in probability, and ``$\stackrel{d}{\longrightarrow}$'' denotes  convergence in distribution.

%and $J_n = I_n - \frac{1}{n}l_nl_n'$.

%$\mathcal{F}_{i,t-1}$ denotes the $\sigma$-field generated by $\epsilon_{i,t-1},\epsilon_{i,t-2},...$ with $\epsilon_{it}$'s being i.i.d. variables.

\section{The model and its estimation method}\label{section_model}

\subsection{The panel ARMA--GARCH model}
Given a panel of observations $\{(y_{it},x_{it})\}$, our panel ARMA--GARCH model is defined as
\begin{subequations}\label{equ_model}
\begin{align}\label{equ_model_ARMA}
y_{it} &= \mu_{0i} + x_{it}'\beta_0 + \sum_{p=1}^{P}\phi_{0p}y_{i,t-p}  + \sum_{q=1}^{Q}\psi_{0q}u_{i,t-q}+u_{it},\\  \label{equ_model_GARCH}
u_{it} &=\sqrt{h_{it}}\epsilon_{it} \text{ with } h_{it} =  \varpi_{0i} + \sum_{l=1}^{L}\tau_{0l} u_{i,t-l}^2 + \sum_{k=1}^{K} \nu_{0k} h_{i,t-k},
\end{align}
\end{subequations}
for $i=1,...,N$ and $t=1,...,T$, where $\mu_{0i}$ and $\varpi_{0i}$ are the unobservable fixed effects, $\beta_0$ is a
$D_x$-dimensional vector of unknown parameters,
$\phi_{0p}$ and $\psi_{0q}$ are the ARMA parameters, $\tau_{0l}$ and $\nu_{0k}$ are the GARCH parameters,
and $\{\epsilon_{it}\}$ is a sequence of i.i.d. errors with mean zero and variance one. Here, we assume $\varpi_{0i}>0$, $\tau_{0l}\geq 0$, and $\nu_{0k}\geq 0$ to ensure $h_{it}>0$, where $h_{it}$ is the conditional variance of $y_{it}$ given the $\sigma$-field generated by $\{\epsilon_{i,t_*}\}_{t_*\leq t-1}$. Clearly, the panel ARMA--GARCH model contains two parts:
the panel ARMA specification with orders $P$ and $Q$ in (\ref{equ_model_ARMA}) and the panel GARCH specification with orders $L$ and $K$ in (\ref{equ_model_GARCH}), where the panel ARMA specification nests the dynamic panel specification in (\ref{model_dynamic_panel}), and the panel GARCH specification is the same as that in \cite{Pakel2011}.

It is worth noting that the panel ARMA--GARCH model serves as a bridge, connecting the dynamic panel literature with the high-dimensional time series literature. Compared with the high-dimensional ARMA or GARCH model, the panel ARMA--GARCH model has a crucial panel constraint that all of $y_{it}$ share the same ARMA and GARCH parameters cross-sectionally. This panel constraint is common in the dynamic panel literature. It allows us to use the entire panel observations to estimate these shared parameters, so the resulting estimators can have the asymptotic normality with the convergence rate $(NT)^{-1/2}$.
Like most of dynamic panel models, the panel ARMA--GARCH model also inherits their another common feature that the dynamics of
$y_{it}$ only depends on their own lagged values but not on the lagged values of other $y_{jt}$.
This cross-sectional independence feature is not often assumed in the high-dimensional ARMA or GARCH model, although
it could help to avoid over-parameterization, improve prediction, and match an empirical finding that $y_{it}$ relies more on its own past than it does on the past of other $y_{jt}$ (\citealp{EK1995}).
To capture the cross-section dependence, we could follow the common way to add the spatial or network structure in the panel ARMA--GARCH model as done in \cite{Zhu2017}, \cite{Kuersteiner2020}, and \cite{Zhou2020}. It appears that our estimation method and related technical treatments can be extended to these spatial and network cases, which are not investigated in detail for ease of exposition.

%Moreover, the spatial or network matrix is often used to capture the cross-section dependence in the panel literature, see e.g., \cite{Zhu2017}, \cite{Kuersteiner2020}, and \cite{Zhou2020}. It appears that our estimation method can be extended to the spatial or network cases.

\subsection{The two-step estimation method}\label{subsection_estimation}

Due to the presence of $\mu_{0i}$ and $\varpi_{0i}$, the estimation of our panel ARMA--GARCH model has the incidental parameter problem. To solve this problem, we design a two-step estimation method to estimate panel ARMA specification and panel GARCH specification stepwisely.

In the first step, we estimate the panel ARMA specification in (\ref{equ_model_ARMA}) by using the LS estimation method.
Let $\theta_1 =  (\mu',\lambda') \in \Theta_1$ be the parameter vector in (\ref{equ_model_ARMA})
and $\theta_{01} =  (\mu_{0}',\lambda_{0}') \in \Theta_1$ be its true value, where $\Theta_1=\Theta_{\mu}\times \Theta_{\lambda}\subset \mathcal{R}^{N+D_x+P+Q}$ is the parameter space of $\theta_1$, $\mu=(\mu_1,...,\mu_{N})'\in\Theta_{\mu}\subset \mathcal{R}^{N}$, $\lambda=(\beta',\phi', \psi')'\in\Theta_{\lambda}\subset\mathcal{R}^{D_x+P+Q}$ with $\beta\in\mathcal{R}^{D_x}$, $\phi = (\phi_1,...,\phi_P)'\in\mathcal{R}^{P}$, and $\psi =  (\psi_1,...,\psi_Q)'\in\mathcal{R}^{Q}$, and $\mu_{0}$ and $\lambda_{0}$ are defined analogously.
For $u_{it}$, its parametric form is $u_{it,\theta_1}$ defined iteratively by
$u_{it,\theta_1}=y_{it}-\mu_i-x_{it}'\beta-\sum_{p=1}^{P}\phi_{p}y_{i,t-p}-\sum_{q=1}^{Q}\psi_{q}u_{i,t-q,\theta_1}$.
By construction, we have $u_{it}=u_{it,\theta_{01}}$. However, $u_{it,\theta_1}$ is computationally infeasible due to some unobservable initial values.
Therefore, we have to consider $\hat{u}_{it,\theta_1}$ (i.e., the computationally feasible version of $u_{it,\theta_1}$) defined iteratively by
$\hat{u}_{it,\theta_1}=y_{it}-\mu_i-x_{it}'\beta-\sum_{p=1}^{P}\phi_{p}y_{i,t-p}-\sum_{q=1}^{Q}\psi_{q}\hat{u}_{i,t-q,\theta_1}$ for $i=1,...,N$ and $t=1,...,T$, with the initial values $y_{i,1-P}=\cdots=y_{i,0}=\hat{u}_{i,1-Q,\theta_1}=\cdots=\hat{u}_{i,0,\theta_1}=0$.
Let $\hat{U}_{\theta_1} = (\hat{u}_{1,\theta_1}',...,\hat{u}_{N,\theta_1}')'\in\mathcal{R}^{NT}$ with $\hat{u}_{i,\theta_1}=(\hat{u}_{i1,\theta_1},...,\hat{u}_{iT,\theta_1})'\in\mathcal{R}^{T}$. Then,
$\hat{U}_{\theta_1}$ satisfies the following equation:
\begin{equation}\label{equ_model_ARMA_tran}
(I_N\otimes A_{\phi}) Y =  (I_N\otimes l_T)\mu  +  X\beta  + (I_N \otimes B_{\psi}) \hat{U}_{\theta_1},
\end{equation}
where $Y=(y_1',...,y_N')'\in\mathcal{R}^{NT}$ with $y_i=(y_{i1},...,y_{iT})'\in\mathcal{R}^{T}$, $X=(x_1',...,x_N')'\in\mathcal{R}^{NT\times D_x}$ with $x_i=(x_{i1},...,x_{iT})\in\mathcal{R}^{D_x\times T}$, $\mu = (\mu_1,...,\mu_N)'\in\mathcal{R}^{N}$, and $A_{\phi}$ and $B_{\psi}$ are two $T\times T$ invertible matrices defined by
\begin{equation*}
A_{\phi}=
\left[
\begin{array}{cccccc}
1& & & & &\\
- \phi_1&\ddots & & & &\\
\vdots&\ddots &\ddots& & &\\
- \phi_P&\ddots & \ddots&\ddots & &\\
& \ddots&\ddots &\ddots &\ddots &\\
& &- \phi_P& ...& -\phi_1 & 1\\
\end{array}\right] \mbox{ and }
B_{\psi}=
\left[
\begin{array}{cccccc}
1& & & & &\\
\psi_1&\ddots & & & &\\
\vdots&\ddots &\ddots& & &\\
\psi_Q&\ddots & \ddots&\ddots & &\\
& \ddots&\ddots &\ddots &\ddots &\\
& &\psi_Q& ...&\psi_1 & 1\\
\end{array}\right].
\end{equation*}
Clearly,  $\hat{U}_{\theta_1}$ is the computationally feasible version of $U_{\theta_1}$, where
$U_{\theta_1} = (u_{1,\theta_1}',...,u_{N,\theta_1}')'\in\mathcal{R}^{NT}$ with $u_{i,\theta_1}=(u_{i1,\theta_1},...,u_{iT,\theta_1})'\in\mathcal{R}^{T}$.

From (\ref{equ_model_ARMA_tran}), we know that $\hat{U}_{\theta_1}$ has the form
\begin{align}\label{equ_UVC}
\hat{U}_{\theta_1} = (I_N \otimes B_{\psi}^{-1})(V_{\phi,\beta} - (I_N\otimes l_T)\mu) \mbox{ with } V_{\phi,\beta} = (I_N\otimes A_{\phi}) Y - X\beta.
\end{align}
Then, our objective function for the LS estimation is defined as $\hat{Q}_{\theta_1}=\hat{U}_{\theta_1}' \hat{U}_{\theta_1}$. By solving the equations $\partial \hat{Q}_{\theta_1}/\partial \mu = 0$, the solution of $\mu$ for any given $\lambda$ is
\begin{equation}\label{equ_hat_mu}
\hat\mu_{\lambda} = (I_N \otimes (l_T'\Sigma_{\psi}^{-1}l_T)^{-1}l_T'\Sigma_{\psi}^{-1}) V_{\phi,\beta},
\end{equation}
where $\Sigma_{\psi} = B_{\psi}B_{\psi}'$. Furthermore, by replacing $\mu$ with $\hat\mu_{\lambda}$ in $\hat{Q}_{\theta_1}$, we get the concentrated objective function $\hat{Q}_{\lambda}=V_{\phi,\beta}'(I_N \otimes (\Sigma^{-1}_{\psi} - C_{\psi})) V_{\phi,\beta}$,
where $C_{\psi} = (l_T'\Sigma_{\psi}^{-1}l_T)^{-1} \Sigma_{\psi}^{-1}l_Tl_T'\Sigma_{\psi}^{-1}$.
Based on $\hat{Q}_{\lambda}$, we define the LS estimator of $\lambda_0$ as follows:
\begin{equation}\label{equ_hat_lambda}
\hat\lambda = (\hat\beta',\hat\phi', \hat\psi')' =\arg\min_{\lambda\in\Theta_{\lambda}} \hat{Q}_{\lambda},
\end{equation}
where $\hat\phi=(\hat\phi_1,...,\hat\phi_P)'$ and $\hat\psi=(\hat\psi_1,...,\hat\psi_Q)'$. Substituting $\lambda$ with $\hat\lambda$ in (\ref{equ_hat_mu}), we obtain $\hat\mu_{\hat\lambda}$, which is the LS estimator of $\mu_0$.
In sum, our LS estimator of $\theta_{01}$ is $\hat\theta_1$, where $\hat\theta_1=(\hat\mu_{\hat\lambda}', \hat\lambda')'$.

In the second step, we estimate the panel GARCH specification in (\ref{equ_model_GARCH}) by using the variance-targeting (VT) method (\citealp{Francq2011}).
Let $\theta_2 =  (\varpi',\zeta') \in \Theta_2$ be the parameter vector in (\ref{equ_model_GARCH})
and $\theta_{02} =  (\varpi_{0}',\zeta_{0}') \in \Theta_2$ be its true value, where $\Theta_2=\Theta_{\varpi}\times \Theta_{\zeta}\subset\mathcal{R}^{N+L+K}$, $\varpi=(\varpi_1,...,\varpi_{N})'\in\Theta_{\varpi}\subset \mathcal{R}^{N}$, $\zeta=(\tau',\nu')'\in\Theta_{\zeta}\subset\mathcal{R}^{L+K}$ with $\tau = (\tau_1,...,\tau_L)'\in\mathcal{R}^{L}$ and $\nu =  (\nu_1,...,\nu_K)'\in\mathcal{R}^{K}$, and $\varpi_{0}$ and $\zeta_{0}$ are defined analogously.
To facilitate the VT method, we assume $\sum_{l=1}^{L}\tau_{0l} + \sum_{k=1}^{K} \nu_{0k}<1$, which implies that
$\omega_{0i}:={\rm E} (u_{it}^2) <\infty$ (\citealp{Bollerslev1986}). Then, we have $\varpi_{0i}=\omega_{0i}\big(1-  \sum_{l=1}^{L}\tau_{0l}- \sum_{k=1}^{K} \nu_{0k} \big)$, so we can re-parameterize (\ref{equ_model_GARCH}) as
\begin{equation} \label{equ_model_GARCH_VT}
u_{it} =\sqrt{h_{it}}\epsilon_{it} \text{ with } h_{it} =  \omega_{0i}\Big(1-  \sum_{l=1}^{L}\tau_{0l}- \sum_{k=1}^{K} \nu_{0k} \Big) + \sum_{l=1}^{L}\tau_{0l} u_{i,t-l}^2 + \sum_{k=1}^{K} \nu_{0k} h_{i,t-k}.
\end{equation}
For $h_{it}$ in (\ref{equ_model_GARCH_VT}), it has the parametric form $h_{it,\zeta,\omega_i}$ defined iteratively
by
\begin{equation}\label{equ_h_it_zeta_omegai}
h_{it,\zeta,\omega_i} = \omega_i\Big(1-  \sum_{l=1}^{L}\tau_l- \sum_{k=1}^{K} \nu_k \Big) + \sum_{l=1}^{L}\tau_l u_{i,t-l}^2 + \sum_{k=1}^{K} \nu_k h_{i,t-k,\zeta,\omega_i}
\end{equation}
for $\omega_i>0$. Clearly, $h_{it}=h_{it,\zeta_0,\omega_{0i}}$. By assuming $\epsilon_{it}\stackrel{\text{i.i.d.}}{\sim} N(0, 1)$ in (\ref{equ_model_GARCH_VT}), the log-likelihood function of $\{u_{it}\}$ (ignoring constants) is
\begin{equation}\label{objective_QMLE}
L_{\zeta,\omega}=-\frac{1}{2}\sum_{i=1}^{N}\sum_{t=1}^{T} \Big[\log(h_{it,\zeta,\omega_i}) + \frac{u_{it}^2}{h_{it,\zeta,\omega_i}}\Big],
\end{equation}
where $\omega=(\omega_1,...,\omega_{N})'\in \mathcal{R}^{N}$. As $\{u_{it}\}_{t=1-L}^{T}$ and $\{h_{it,\zeta,\omega_i}\}_{t=1-K}^{0}$ are  unobservable, we have to consider $\hat L_{\zeta,\omega}$ (i.e.,
the computationally feasible version of $L_{\zeta,\omega}$) defined by
\begin{equation*}\label{objective_QMLE_feasible}
\hat{L}_{\zeta,\omega}=-\frac{1}{2}\sum_{i=1}^{N}\sum_{t=1}^{T} \Big[\log(\hat{h}_{it,\zeta,\omega_i}) + \frac{\hat{u}_{it}^2}{\hat{h}_{it,\zeta,\omega_i}}\Big],
\end{equation*}
where $\hat{h}_{it,\zeta,\omega_i}$ (i.e., the computationally feasible version of $h_{it,\zeta,\omega_i}$) is defined iteratively
by
$$\hat{h}_{it,\zeta,\omega_i} = \omega_i\Big(1-  \sum_{l=1}^{L}\tau_l- \sum_{k=1}^{K} \nu_k \Big) + \sum_{l=1}^{L}\tau_l \hat{u}_{i,t-l}^2 + \sum_{k=1}^{K} \nu_k \hat{h}_{i,t-k,\zeta,\omega_i},$$
with the initial values
$\hat{u}_{i,1-L}=\cdots=\hat{u}_{i0}=0$ and  $\hat{h}_{i,1-K,\zeta,\omega_i}=\cdots=\hat{h}_{i,0,\zeta,\omega_i}=c_{h}$. Here,
$\hat u_{it}:=\hat u_{it,\hat{\theta}_1}$ is the residual computed from $\hat U_{\hat\theta_1}=(\hat u_{1,\hat{\theta}_1}',...,\hat u_{N,\hat{\theta}_1}')'$ with $\hat u_{i,\hat{\theta}_1}  = (\hat u_{i1,\hat{\theta}_1},...,\hat u_{iT,\hat{\theta}_1})'$, and $c_{h}>0$ is a given constant.

Like $\varpi$, the presence of $\omega$ in $\hat L_{\zeta,\omega}$ causes the incidental parameter problem, but we cannot concentrate out
$\omega$ since the equation $\partial \hat L_{\zeta,\omega}/\partial \omega=0$ does not deliver a closed-form solution of $\omega$. To circumvent this deficiency, we simply estimate $\omega_0$ by
\begin{equation*}\label{equ_hat_omega}
\hat\omega=(\hat\omega_1,...,\hat\omega_N)' \mbox{ with }\hat \omega_i = \frac{1}{T}\sum_{t=1}^{T} \hat u_{it}^2,
\end{equation*}
in view of the fact that $\omega_{0i}$ is the second moment of $u_{it}$. After replacing $\omega$ with  $\hat\omega$ in $\hat L_{\zeta,\omega}$,
we further estimate $\zeta_0$ by the VT-QML estimator
\begin{equation}\label{equ_hat_zeta}
\hat\zeta = (\hat\tau', \hat\nu')'= \arg\max_{\zeta\in\Theta_{\zeta}} \hat L_{\zeta,\hat\omega},
\end{equation}
where $\hat\tau=(\hat\tau_1,...,\hat\tau_L)'$ and $\hat\nu=(\hat\nu_1,...,\hat\nu_K)'$. Based on $\hat\omega$ and $\hat\zeta$, we now estimate $\varpi_0$ by
\begin{equation*}\label{equ_hat_varpi}
\hat\varpi=(\hat\varpi_1,...,\hat\varpi_N)' \mbox{ with }\hat\varpi_i=\hat\omega_i\Big(1-  \sum_{l=1}^{L}\hat\tau_l- \sum_{k=1}^{K} \hat\nu_k \Big).
\end{equation*}
To sum up, our VT-QML estimator of $\theta_{02}$ is $\hat\theta_2$, where $\hat\theta_2=(\hat\varpi',\hat\zeta')'$.

We should highlight that the classical QML estimation method estimates the ARMA--GARCH model jointly instead of stepwisely (\citealp{FZ2004}). Here, our main reason to use the two-step estimation method is to tackle the incidental parameter problem, so that $\mu_i$ can be concentrated out at step one and $\varpi_i$ can be re-parameterized out at step two. Clearly, the joint estimation method does not allow us to achieve this goal.

\section{The asymptotics of linear-quadratic form}\label{section_LQ}

To establish the asymptotic theory of our proposed estimators, we need some new asymptotics for the following linear-quadratic form:
\begin{equation}\label{equ_VMV_bV}
\mathcal{LQ}=V' M V + b' V,
\end{equation}
where $V=(v_1',...,v_{N}')'$ is an $NT$-dimensional vector of variables with $v_i=(v_{i1},...,v_{iT})'\in\mathcal{R}^{T}$, $M$ is an $NT\times NT$ non-stochastic block weight matrix with $(i,j)$-th block $M_{ij}=[m_{ij,tt^*}]\in\mathcal{R}^{T\times T}$,
and $b=(b_1',...,b_N')'$ is an $NT$-dimensional non-stochastic weight vector with $b_i=(b_{i1},...,$ $b_{iT})'\in\mathcal{R}^{T}$. Note that the elements of $M$ and $b$ are allowed to depend on $N$ and $T$, and we have suppressed their subscripts $N$ and $T$ for ease of presentation.

When $N=1$ and $b=0$, the linear-quadratic form $\mathcal{LQ}$ reduces to the quadratic form $v_1'M_{11}v_1$. Under the assumption that $m_{11,tt^*}$ is a function of $|t-t^*|$, \cite{Whittle1964}, \cite{Giraitis1997,Giraitis1998}, and \cite{Hsing2004} derive the CLT for the quadratic form $v_1'M_{11}v_1$, where $\{v_{1t}\}$ is an i.i.d. or martingale difference sequence. However, the above assumption on $m_{11,tt^*}$ is restrictive. For example, it rules out the quadratic form with $M_{11}=I_T - l_Tl_T'/T$, which can appear in our objective function $\hat{Q}_{\lambda}$ (through $\Sigma_{\psi}^{-1} - C_{\psi}$) in (\ref{equ_hat_lambda}) and has to be tackled. \cite{Wu2007} relieves this assumption by posing the assumption
$T^{-1}\sum_{t = 1}^{T- t^*} m_{11, t,t+t^*}^2 = o(1)$ for any given $0\le t^* < T$ and some other assumptions on $M_{11}$, which remain inapplicable to the quadratic form with $\Sigma_{\psi}^{-1} - C_{\psi}$.
\cite{Giraitis2017} provides some assumptions about the Euclidean and spectral norms of matrix $M_{11}$, which is difficult to check for matrices appearing in our proofs for the proposed estimators.

When $T=1$ and $b\not=0$, \cite{Kelejian2001} establishes the CLT for the linear-quadratic form $\mathcal{LQ}$, provided that $M$ is uniformly bounded in row and column and $\{v_{i1}\}$ is an i.i.d. sequence.
%See also some similar asymptotic results in \cite{Yu2008} and \cite{Qu2017}.
Although the condition of $M$ is desirable, the proof technique in \cite{Kelejian2001} is not transferable to the case
of $T>1$, under which there is temporal dependence in each $v_i$.

Below, we provide the asymptotics for the linear-quadratic form $\mathcal{LQ}$, based on the following general conditions on $M$, $b$, and $V$ in (\ref{equ_VMV_bV}).

\begin{cond}\label{cond_uniformbound}
(i) $\sup_{i,t}\sum_{j=1}^{N}\sum_{t^* = 1}^{T} |m_{ij,tt^*}|<\infty$ and $\sup_{j,t^*}\sum_{i=1}^{N}\sum_{t=1}^{T}|m_{ij,tt^*}|<\infty$. (ii) $\sup_{i,t} b_{it}^2<\infty$.
\end{cond}

\begin{cond}\label{cond_v}
(i) $\{v_i\}$ is an independent sequence. (ii) For each $i$, $\{v_{it}\}$ is a strictly stationary and uncorrelated sequence, and $v_{it}$ is $L_{4+\delta}$-bounded for some $\delta>0$ with ${\rm E}(v_{it})=0$. (iii) For each $t$, the following moment conditions hold:
\begin{align*}
|a_{it}| < &\infty \text{ with } a_{it} = {\sum}_{t^* = 1, t^* \neq t}^{T} (m_{ii,tt}m_{ii,t^*t^*} + m_{ii,tt^*}^2 + m_{ii,tt^*}m_{ii,t^*t})\varsigma_{i,t,t^*},\\
|e_{it}| < &\infty \text{ with } e_{it} = {\sum}_{t^* = 1, t^* \neq t}^{T}{\sum}_{t^{\sharp} = 1, t^{\sharp} \neq t, t^*}^{T} (m_{ii,tt} m_{ii,t^*t^{\sharp}} + m_{ii,tt^*} m_{ii,tt^{\sharp}}+  m_{ii,tt^*} m_{ii,t^{\sharp}t}\\
&  \qquad\qquad \qquad\qquad\qquad\qquad\qquad\quad\,\, + m_{ii,t^*t} m_{ii,t^{\sharp}t} + m_{ii,t^*t} m_{ii,tt^{\sharp}}) \vartheta_{i,t,t^*,t^{\sharp}},\\
|f_{it}| < &\infty \text{ with } f_{it} = {\sum}_{t^* = 1, t^* \neq t}^{T}(m_{ii,tt}m_{ii,tt^*} +m_{ii,tt}m_{ii,t^*t}) \varrho_{i,t,t^*},\\
|g_{it}| < &\infty \text{ with } g_{it} = {\sum}_{t^* = 1, t^* \neq t}^{T} (m_{ii,tt}b_{it^*} + m_{ii,tt^*}b_{it} + m_{ii,t^*t}b_{it}) \pi_{i,t,t^*},
\end{align*}
where $\varsigma_{i,t,t^*} = {\rm Cov}(v_{it}^2, v_{it^*}^2)$, $\vartheta_{i,t,t^*,t^{\sharp}} = {\rm E}(v_{it}^2 v_{it^*}v_{it^{\sharp}})$, $\varrho_{i,t,t^*} = {\rm E}(v_{it}^3 v_{it^*})$,  and $\pi_{i,t,t^*} = {\rm E}(v_{it}^2 v_{it^*})$.
\end{cond}

We offer some remarks on the above two conditions. Condition \ref{cond_uniformbound}(i) requires that
$M$ is uniformly bounded in row and column, and Condition \ref{cond_uniformbound}(ii) holds when elements in $b_i$ are uniformly bounded.
This condition is similar to the one in \cite{Kelejian2001}, and it does not need to assume that $|m_{ij,tt^*}|$ is a function of
$|t-t^*|$. Condition \ref{cond_v}(i)--(ii) assume that $\{v_{it}\}$ are i.i.d. across $i$ and strictly stationary in $t$ for each $i$, and they are also uncorrelated for each $i$. These settings are common in dynamic panel models (see \cite{Hahn2002}), and if each $v_i$ is viewed as a block of $v$, they essentially allow $v$ to be block-independent with certain temporal dependence within the block. Meanwhile, the stationarity and $L_{4+\delta}$-boundedness condition in Condition \ref{cond_v}(ii) ensures the existence of
$\varsigma_{i,t,t^*}$, $\vartheta_{i,t,t^*,t^{\sharp}}$, $\varrho_{i,t,t^*}$, and $\pi_{i,t,t^*}$ in Condition \ref{cond_v}(iii). The moment conditions in
Condition \ref{cond_v}(iii) are regular and they are satisfied if Conditions \ref{cond_uniformbound}--\ref{cond_v}(ii) hold,
\begin{equation}\label{checking_cond_1}
\sum_{t^* = 1, t^* \neq t}^{T}|m_{ii,t^*t^*}\varsigma_{i,t,t^*}| <\infty,  \,\mbox{ and }\,
\sum_{t^* = 1, t^* \neq t}^{T}{\sum}_{t^{\sharp} = 1, t^{\sharp} \neq t, t^*}^{T} | m_{ii,t^*t^{\sharp}}\vartheta_{i,t,t^*,t^{\sharp}}|  <\infty.
\end{equation}
The sufficient conditions in (\ref{checking_cond_1}) are mild, and they hold for many time series specifications. For example,
when $v_{it}$ follows the GARCH specification with a finite fourth moment for each $i$, the conditions in (\ref{checking_cond_1}) are valid, since
$\varsigma_{i,t,t^*}=O(c_0^{|t-t^*|})$ and $\vartheta_{i,t,t^*,t^{\sharp}} = O(c_0^{|t - \max\{t^*,t^{\sharp}\}|})$ for some $c_0\in(0, 1)$ (\citealp{Francq2019}). It is worth noting that
our Condition \ref{cond_v}(iii) is different from the high-order cumulative summability conditions on $\{v_{it}\}$ for each $i$ in \cite{Hahn2002}, and it is easy to check as demonstrated above. For the validity of high-order cumulative summability conditions, some mixing conditions are generally needed but they can be difficult to verify for many time series specifications.

To present the asymptotics of $\mathcal{LQ}$, we first give the formulas of its mean and variance:
\begin{align*} %\label{equ_sigma_LQ}
\mu_{\mathcal{LQ}}&={\rm E}(\mathcal{LQ})=\sum_{i=1}^{N} \sigma_i^2\text{tr}(M_{ii}),\\
\sigma_{\mathcal{LQ}}^2&={\rm Var}(\mathcal{LQ})=\sum_{i=1}^{N}\sum_{t=1}^{T}(\varrho_i - 3\sigma_i^4)m_{ii,tt}^2 +  \sum_{i=1}^{N}\sum_{t=1}^{T}  (a_{it} + 2e_{it} +  2 f_{it} + 2g_{it} +  2 \pi_i b_{i,t} m_{ii,tt}) \\
&\quad\quad\quad\quad\quad\quad  + \sum_{i=1}^{N} \sum_{j=1}^{N} \sigma_i^2\sigma_j^2[\text{tr}(M_{ij}M_{ij}') + \text{tr}(M_{ij}M_{ji})]  +  \sum_{i=1}^{N} \sigma_i^2 b_i'b_i,
\end{align*}
where $\sigma_i^2={\rm E}(v_{it}^2)$, $ \pi_i = {\rm E}(v^3_{it})$, $\varrho_i = {\rm E}(v_{it}^4)$,
and $a_{it}$, $e_{it}$, $f_{it}$, and $g_{it}$ are defined as in Condition \ref{cond_v}. Next, we show the convergence rate of
$\mathcal{LQ}$ is $(NT)^{-1/2}$.

\begin{thm}\label{thm_LLN_uncorrelated}
Suppose that Conditions \ref{cond_uniformbound}--\ref{cond_v} hold. If either $N$ or $T$ is large,
\begin{equation*}
\mathcal{LQ}- \mu_{\mathcal{LQ}}= O_p(\sqrt{NT}).
\end{equation*}
\end{thm}

Moreover, to establish the CLT of $\mathcal{LQ}$, we need the condition below:

\begin{cond}\label{cond_MD}
(i) For each $i$, $\{v_{it}\}$ is ergodic and a martingale difference sequence with respect to the filtration
$\mathcal{G}_{i,-\infty}^{t}$, where $\mathcal{G}_{i,a}^{b}$ is a $\sigma$-field generated by $\{v_{it^*}\}_{t^{*}=a}^{b}$.

(ii) Either the conditional variance $\sigma_i^2 = {\rm E}(v_{it}^2| v_{i,t-1},...)$ for all $t$, or the matrix $M_{ii}$ satisfies (a)
$T^{-1}\sum_{t=1}^{T}m_{ii,tt}^2 = o(1)$; (b) $T^{-1}\sum_{t=t^*+2}^{T}|m_{ii,t,t-t^*} -m_{ii,t-1,t-t^*-1}| = o(1)$ with any $1\le t^*< T -1 $; and (c) $T^{-1}\sum_{t=1}^{T}\sum_{t^*=1, |t-t^*|\ge \chi}^{T}m_{ii, tt^*}^2 = o(1)$ as $T, \chi \to \infty$.
\end{cond}

%\begin{cond}\label{cond_mixing}
%For each $i$, $\{v_{it}\}$ is $\alpha$-mixing with the mixing coefficient $\alpha_{i,m}$, where $\alpha_{i,m}= \sup_t\sup_{A\in \mathcal{G}_{i,-\infty}^t, B\in \mathcal{G}_{i,t+m}^{\infty}}|P(A B) - P(A)P(B)|$ satisfying
%$\sup_i \sum_{m=0}^{\infty}\alpha_{i,m}^{\delta/(4+\delta)}<\infty$.
%\end{cond}

Condition \ref{cond_MD} poses some regular conditions on $\{v_{it}\}$ for each $i$.
Specifically, Condition \ref{cond_MD}(i) holds when $v_{it}$ has the GARCH specification.
Condition \ref{cond_MD}(ii) is made to handle $\{ v_{it}^2 - \sigma^2_i\}$,
which is a non-martingale difference sequence with respect to $\mathcal{G}_{i,-\infty}^{t}$.
When $\{v_{it}\}$ is an i.i.d. sequence across $t$, we have
${\rm E}(v_{it}^2| v_{i,t-1},...)={\rm E}(v_{it}^2)$ for all $t$, so Condition \ref{cond_MD}(ii) holds.
In general, when $\{v_{it}\}$ has certain temporal dependence structure such as GARCH,  ${\rm E}(v_{it}^2| v_{i,t-1},...)\not={\rm E}(v_{it}^2)$ for all $t$. Then, we can verify Condition \ref{cond_MD}(ii) by checking these matrix assumptions
in parts (a)--(c), which hold in our following theoretical analysis for the proposed estimators. For similar matrix assumptions, one can refer to \cite{Wu2007} and \cite{Giraitis2017}.
%Although Condition \ref{cond_MD} is sufficient for our theoretical analysis, we provide
%Condition \ref{cond_mixing} for the non-martingale difference but $\alpha$-mixing sequence $\{v_{it}\}$.
Now, we give the CLT for $\mathcal{LQ}$.

%\cite{Yu1990},

\begin{thm}\label{thm_LLN_CLT_MD}
Suppose that (i) Conditions \ref{cond_uniformbound}--\ref{cond_MD} hold; and (ii) $(NT)^{-1}\sigma_{\mathcal{LQ}}^2\ge c$ for some $c>0$. If either $N$ or $T$ is large,
$$\frac{\mathcal{LQ}- \mu_{\mathcal{LQ}}}{\sigma_{\mathcal{LQ}}} \stackrel{\text{d}}{\longrightarrow}N(0,1).$$
\end{thm}

%\begin{thm}\label{thm_LLN_CLT_mixing}
%	Suppose Conditions \ref{cond_uniformbound}--\ref{cond_v} and \ref{cond_mixing} hold, and $\frac{1}{NT}\sigma_{\mathcal{LQ}}^2\ge c$ for some $c>0$. If either $N$ or $T$ is large, then
%	$$\frac{\mathcal{LQ}}{\sigma_{\mathcal{LQ}}} \stackrel{\text{d}}{\longrightarrow}N(0,1).$$
%\end{thm}

Since the presence of exogenous variables $X$ in (\ref{equ_model_ARMA_tran}) can make $b$ stochastic in our theoretical analysis, we further give a corollary to allow for
the exogenous $b$.

%\begin{cond}\label{cond_b_random}
%	(i) $b = (b_1',...,b_N')'$ is an $NT\times 1$-dimensional stochastic vectors with $b_i = [b_{it}]$. $b_{it}$'s are independent with $v_{it}$, and  $\sup_{i,t} {\rm E}(b_{it}^2) <\infty$.
%	
%	(ii) $f_{it}$ and $\sigma_{\mathcal{LQ}}^2$ are redefined by replacing ${\rm E}(b_{it})$ and ${\rm E}(b_{it}^2)$ as $b_{it}$ and $b_{it}^2$.
%\end{cond}

\begin{cor}\label{cor_Q}
Suppose that (i) Conditions \ref{cond_uniformbound}--\ref{cond_v} hold with $b_{it}$ replaced by ${\rm E}(b_{it})$; (ii) Condition \ref{cond_MD} holds; (iii) $(NT)^{-1}\sigma_{\mathcal{LQ}}^2\ge c$ for some $c>0$. If either $N$ or $T$ is large,
$$\frac{\mathcal{LQ}- \mu_{\mathcal{LQ}}}{\sigma_{\mathcal{LQ}}} \stackrel{\text{d}}{\longrightarrow}N(0,1),$$
where $\sigma_{\mathcal{LQ}}$ is defined with $b_i$ and $b_{i,t}$ replaced by ${\rm E}(b_i)$ and ${\rm E}(b_{i,t})$.
\end{cor}

\section{The asymptotics of all proposed estimators}\label{section_asymptotics}

\subsection{The technical assumptions}
Let $\theta=(\theta_1',\theta_2')'\in\Theta$ be an $\mathcal{S}$-dimensional vector of unknown parameters with the true value $\theta_0=(\theta_{01}',\theta_{02}')'\in\Theta$ in (\ref{equ_model_ARMA})--(\ref{equ_model_GARCH}), where $\Theta=\Theta_1\times\Theta_2\subset \mathcal{R}^{\mathcal{S}}$ is the parameter space of $\theta$, and $\mathcal{S}=2N+D_x+P+Q+L+K$. Denote $\phi(z) = 1 - \sum_{p=1}^{P}\phi_pz^p$, $\psi(z) = 1 + \sum_{q=1}^{Q}\psi_qz^q$, $\tau(z) = \sum_{l=1}^{L}\tau_lz^l$, and $\nu(z) = 1- \sum_{k=1}^{K} \nu_kz^k$. To derive the asymptotics of our proposed estimators, we make the following assumptions:

\begin{ass}\label{ass_parameters}
(i) $\Theta$ is compact and $\theta_0$ is an interior point of $\Theta$.

(ii) For each $\lambda\in \Theta_{\lambda}$, 	
both $\phi(z)$ and $\psi(z)$ have no common root with $\phi_P \neq 0$ and $\psi_Q \neq 0$; moreover, $\phi(z)\neq 0$ and $\psi(z)\neq 0$ when $|z|\le1$, $\sum_{p=1}^{P}|\phi_{p}|<\infty$,
and $\sum_{q=1}^{Q}|\psi_{q}|<\infty$.

(iii) For each $\zeta \in \Theta_{\zeta}$, $\tau(z)$ and $\nu(z)$ have no common root with $\tau_L +\nu_K \neq 0$,
and $\sum_{l=1}^{L}\tau_l + \sum_{k=1}^{K} \nu_k <1$.
\end{ass}

\begin{ass}\label{ass_x_u_epsilon}
(i) $\{x_{it}\}$  are  non-stochastic and uniformly bounded in $i$ and $t$. Or, they are strictly exogenous, independent across $i$ and $t$,  strictly stationary in $t$ for each $i$, and $L_{4+\delta}$-bounded for some $\delta>0$.

(ii) $\{\epsilon_{it}\}$ are i.i.d. variables with mean zero and variance one, and they are uniformly $L_{4+\delta}$-bounded for some $\delta>0$.
\end{ass}

\begin{ass}\label{ass_TN}
$\lim_{N,T\to \infty} N/T = \rho$, where $0 \le \rho < \infty$.
%$N$ is an increasing function of $T$, and $T$ goes to infinity.
\end{ass}

Assumption \ref{ass_parameters} ensures the stationarity, invertibility, and identifiability of the ARMA--GARCH model for each individual (see, e.g., \cite{Brockwell2002} and \cite{Francq2019}).
Particularly, the condition of $\sum_{l=1}^{L}\tau_l + \sum_{k=1}^{K} \nu_k <1$ is consistent with the finite second moment of $u_{it}$ (i.e., $\sum_{l=1}^{L}\tau_{0l} + \sum_{k=1}^{K} \nu_{0k} <1$), and it ensures the applicability of the VT technique.
Assumption \ref{ass_x_u_epsilon} provides some regular conditions for the exogenous variables and model errors. The temporal independence assumption for the exogenous variable is made to ease our theoretical analysis, and it can be generalized into certain martingale difference or mixing assumptions.
Assumption \ref{ass_TN} is common for dynamic panel models; see, for example, \cite{Hahn2002}, \cite{AA2003}, and many others. It states that our asymptotic analysis below is for the case when both $N$ and $T$ diverge to infinity at the same rate.

%the condition of $\sum_{l=1}^{L}\tau_l + \sum_{k=1}^{K} \nu_k <1$ guarantees that
%$u_{it}$ in (\ref{equ_model_GARCH}) has a finite second moment, so that the VT technique is applicable.

%Assumption \ref{ass_x_u_epsilon}(ii) is consistent with the finite second moment of $u_{it}$ (i.e., $\sum_{l=1}^{L}\tau_{0l} + \sum_{k=1}^{K} \nu_{0k} <1$), so that

%All limits are taken under Assumption \ref{ass_TN}, unless stated otherwise.

\subsection{The asymptotics for the panel ARMA specification}\label{subsection_asymptotics_mean}
\subsubsection{The asymptotics of $\hat\lambda$ and $\hat\mu_{\hat\lambda}$}

Denote $\hat D : = \partial \hat Q_{\lambda}/\partial \lambda\big|_{\lambda=\lambda_0}$ and $\hat S := \partial^2 \hat Q_{\lambda}/\partial \lambda \partial \lambda'\big|_{\lambda=\lambda_0}$. Then, we re-write
\begin{align*}%\label{equ_hat_D_hat_S}
\hat D=D+D^{\dag}\,\,\,\mbox{ and }\,\,\,\hat S=S+S^{\dag},
\end{align*}
where  $D^{\dag}=\hat D-D$,  $S^{\dag}=\hat S-S$, and $D$ and $S$ are defined in the same way as $\hat D$ and $\hat S$, respectively, with $\hat U_{\theta_{01}}$ replaced by $U$.
From the proof in the supplementary materials,
\begin{align}
\frac{1}{\sqrt{NT}}\hat D &= \frac{1}{\sqrt{NT}}[D - {\rm E} ( D)] + \frac{1}{\sqrt{NT}}[D^{\dagger} - {\rm E} ( D^{\dagger})] +  \frac{1}{\sqrt{NT}}{\rm E} ( D) + \frac{1}{\sqrt{NT}}{\rm E} ( D^{\dagger}), \nonumber\\
& = O_p(1) + o_p(1) + O\Big(\sqrt{\frac{N}{T}}\Big) + O\Big(\sqrt{\frac{N}{T}}\Big),
\label{equ_Delta_op}
\end{align}
where the first item leads to the asymptotic normality of $\hat\lambda$, the second item is negligible, and the last two items cause the asymptotic bias of $\hat\lambda$. Similarly, from the proof in the supplementary materials, we can show
\begin{align}
\frac{1}{NT} \hat S &= \frac{1}{NT}[ S - {\rm E}( S)] + \frac{1}{NT}[ S^{\dagger} - {\rm E}( S^{\dagger})] + \frac{1}{NT} {\rm E}( S) + \frac{1}{NT}{\rm E}( S^{\dagger}) \nonumber\\
&= o_p(1) +  o_p(1)  + O(1)  +  o_p(1),\label{equ_S_op}
\end{align}
where the third item contributes to the asymptotic variance of $\hat\lambda$, and the rest three items are negligible.

By (\ref{equ_Delta_op})--(\ref{equ_S_op}) and an additional technical assumption below, we can obtain the asymptotics of $\hat\lambda$ in Theorem \ref{thm_lambda_consistent_asymptotic}.

\begin{ass}\label{ass_Gamma_Omega}
Both $\lim \Omega_{1}$ and $\lim \Gamma_{1}$ exist, and  $\lim\Gamma_{1}$ is non-singular, where
\begin{equation*}\label{equ_Delta_Gamma}
%\widetilde{D}_{\lambda_0} = D_{\lambda_0} - {\rm E}(D_{\lambda_0}),\qquad
\Omega_{1} = \frac{1}{NT}{\rm E}\{[D - {\rm E}(D)][D - {\rm E}(D)]'\}\,\,\mbox{ and }\,\,
\Gamma_{1} = - \frac{1}{NT} {\rm E}( S).
\end{equation*}
\end{ass}

\begin{thm}\label{thm_lambda_consistent_asymptotic}
Suppose that Assumptions \ref{ass_parameters}--\ref{ass_TN} hold. Then,
$\hat\lambda\stackrel{\text{p}}{\longrightarrow}\lambda_0$. Moreover, if Assumption \ref{ass_Gamma_Omega} also holds,
\begin{equation*}
\sqrt{NT}\Big(\hat\lambda-\lambda_0 - \frac{1}{T}c_{1} - \frac{1}{T}c_{1}^{\dagger} \Big)  \stackrel{\text{d}}{\longrightarrow} N(0,\Sigma_{1}),
\end{equation*}
where
\begin{equation*}\label{equ_Sigma_ML}
c_{1}=   \frac{\Gamma_{1}^{-1}{\rm E}(D)}{N} = O(1), \,\,\,\,
c_{1}^{\dagger} =   \frac{\Gamma_{1}^{-1}{\rm E}(D^{\dagger})}{N} = O(1),\,\, \mbox{ and }\,\,
\Sigma_{1} = \lim \Gamma_{1}^{-1} \Omega_{1} \Gamma_{1}^{-1}.
\end{equation*}
\end{thm}

\begin{rem}
It is worth noting that the asymptotics of $\hat\lambda$ holds under a general specification of $u_{it}$ rather than GARCH.
To be specific, under Assumptions \ref{ass_parameters}(i)--(ii), \ref{ass_x_u_epsilon}(i), and \ref{ass_TN}--\ref{ass_Gamma_Omega}, the asymptotics in Theorem \ref{thm_lambda_consistent_asymptotic} hold as long as $u_{it}$ satisfies Conditions \ref{cond_v}--\ref{cond_MD}.
\end{rem}

\begin{rem}\label{rem_2}
For saving space, the explicit formulas of ${\rm E}(D)$, ${\rm E}(D^{\dagger})$, $\Gamma_{1}$, and $\Omega_{1}$ are given in Section B.1 of the supplementary materials. Then, under the conditions in Theorem \ref{thm_lambda_consistent_asymptotic}, we can consistently estimate $c_{1}$, $c_{1}^{\dagger}$, and $\Sigma_{1}$ by their plug-in counterparts $\hat{c}_{1}$, $\hat{c}_{1}^{\dagger}$, and $\hat{\Sigma}_{1}$, respectively.
\end{rem}

From Theorem \ref{thm_lambda_consistent_asymptotic}, we find that when $N/T\to 0$, $\hat\lambda$ is $\sqrt{NT}$-consistent and asymptotically
centered normal; when $N/T\to \rho<\infty$ (i.e., both $N$ and $T$ diverge to infinity at the same rate), $\hat\lambda$ is still $\sqrt{NT}$-consistent but asymptotically non-centered normal with mean $\sqrt{\rho}(c_{1} + c_{1}^{\dagger})$.

Notably, the terms ${\rm E}(D)$ and ${\rm E} ( D^{\dagger})$ reflect the influence of the fixed effects and unobservable initial values on the asymptotic bias of $\hat\lambda$, respectively. If $Q=0$ (i.e., (\ref{equ_model_ARMA}) is a panel AR($P$) model), we can show that
${\rm E}(D^{\dagger}) = 0$; therefore, as expected, there is no asymptotic bias resulting from the unobservable initial values in this case. Particularly, if $P=1$ and $Q=0$ (i.e., (\ref{equ_model_ARMA}) is a panel AR(1) model),
our asymptotic normality result in Theorem \ref{thm_lambda_consistent_asymptotic} is consistent to that in \cite{Hahn2002}.
Unlike \cite{Hahn2002} requiring $u_{it}$ to be a mixing sequence, our technical treatment of Theorem \ref{thm_lambda_consistent_asymptotic} works for a more general model and allows $u_{it}$ to be the martingale difference sequence.

Let $\hat\mu_{\hat\lambda,i}$ be the $i$-th entry of $\hat\mu_{\hat\lambda}$.
The following theorem shows that $\hat\mu_{\hat\lambda,i}$ is $\sqrt{T}$-consistent and asymptotically normal without any asymptotic bias.

\begin{thm}\label{thm_mu_eta}
Suppose that Assumptions \ref{ass_parameters}--\ref{ass_Gamma_Omega} hold. Then, for each fixed $i$,
\begin{equation*}
\sqrt{T} \big(\hat\mu_{\hat\lambda,i} - \mu_{0i}\big) \stackrel{\text{d}}{\longrightarrow} N(0, \sigma_{1i}^2),
\end{equation*}
where $\sigma_{1i}^2 =  {\rm E}(u_{it}^2)[\lim T(l_T'\Sigma_{\psi_0}^{-1}l_T)^{-1}]$, and $\Sigma_{\psi}$ is defined as in (\ref{equ_hat_mu}).
\end{thm}

\subsubsection{Bias correction of $\hat\lambda$}
From Theorem \ref{thm_lambda_consistent_asymptotic}, we find that
$\hat\lambda$ has some asymptotic biases when $N/T\not\to 0$.
To achieve a better finite sample performance than $\hat\lambda$, we consider the following analytical bias correction estimator
\begin{equation}\label{equ_lambda_A}
\hat\lambda_A  = \hat\lambda - \frac{1}{T} \hat{c}_{1} - \frac{1}{T} \hat{c}_{1}^{\dagger},
\end{equation}
where $\hat{c}_{1}$ and $\hat{c}_{1}^{\dagger}$ in Remark \ref{rem_2}
are the plug-in estimators of $c_{1}$ and $c_{1}^{\dagger}$, respectively.
Although $\hat\lambda_A$ is expected to have a nice finite sample performance, the calculation of $\hat{c}_{1}$ and $\hat{c}_{1}^{\dagger}$ could become tedious when the orders of the panel ARMA specification are large. Hence, to avoid this computational issue, we follow the idea of
\cite{Dhaene2015} to propose the half-panel Jackknife bias correction estimator
\begin{equation}\label{equ_lambda_J}
\hat\lambda_J = 2 \hat\lambda - \frac{1}{2}(\hat\lambda_1 + \hat\lambda_2),
\end{equation}
where $\hat\lambda_1$ and $\hat\lambda_2$ are defined in the same way as $\hat\lambda$, based on the observations for $t\in \{1,2,..., \lfloor T/2 \rfloor\}$ and those for $t\in \{\lfloor T/2 \rfloor + 1,...,T\}$, respectively. Note that the equation
(\ref{equ_lambda_J}) can be re-written as
$\hat\lambda_J - \lambda_0 = \hat\lambda - \lambda_0 - [(\hat\lambda_1 + \hat\lambda_2)/2 - \hat\lambda]$. Thus, $(\hat\lambda_1 + \hat\lambda_2)/2 - \hat\lambda$ is actually an estimator for the asymptotic bias, and its calculation does not rely on the explicit formulas of the asymptotic bias. The asymptotic normality of $\hat\lambda_A$ and $\hat\lambda_J$ is given as follows:

\begin{thm}\label{thm_lambda_bias_cor}
Suppose that  Assumptions \ref{ass_parameters}--\ref{ass_Gamma_Omega} hold. Then,
\begin{align*}
\sqrt{NT}(\hat\lambda_A - \lambda_0)\stackrel{\text{d}}{\longrightarrow} N(0,\Sigma_{1})\,\,\mbox{ and }\,\,
\sqrt{NT}(\hat\lambda_J - \lambda_0)\stackrel{\text{d}}{\longrightarrow} N(0,\Sigma_{1}).
\end{align*}
%\begin{equation*}
%\sqrt{NT}(\hat\lambda_A - \lambda_0)\stackrel{\text{d}}{\longrightarrow} N(0,\Sigma_{\lambda_0}).
%\end{equation*}
\end{thm}

The above theorem shows that both $\hat\lambda_A$ and $\hat\lambda_J$ are $\sqrt{NT}$-consistent and asymptotically centered normal. Thus, they could have a better finite sample performance than $\hat\lambda$ in terms of bias (see the numerical evidence in Section \ref{section_simulation} below).

\subsection{The asymptotics for the panel GARCH specification}\label{subsection_asymptotics_variance}

\subsubsection{The asymptotics of $\hat\zeta$ and $\hat\varpi$}

The objective function $\hat L_{\zeta, \hat\omega}$ in (\ref{equ_hat_zeta}) involves the unobservable initial values, the estimated fixed effect $\hat\omega$, and the residual $\hat u_{it}$ which
is further based on the ARMA parameter estimator $\hat\lambda$ and the estimated fixed effect $\hat\mu_{\hat\lambda}$. To account for their impact on the asymptotics of $\hat\zeta$, we define $\check{h}_{it,\zeta,\check{\omega}_i}$ and $\tilde{h}_{it,\zeta, \tilde\omega_i}$ recursively by
\begin{align*}
\check{h}_{it,\zeta,\check{\omega}_i} &= \check \omega_i\Big(1-  \sum_{l=1}^{L}\tau_{l}- \sum_{k=1}^{K} \nu_{k} \Big) + \sum_{l=1}^{L}\tau_{l} \check{u}_{i,t-l}^2 + \sum_{k=1}^{K} \nu_{k} \check{h}_{i,t-k,\zeta,\check{\omega}_i},\\
\tilde{h}_{it,\zeta, \tilde\omega_i} &= \tilde \omega_i \Big(1-  \sum_{l=1}^{L}\tau_{l}- \sum_{k=1}^{K} \nu_{k} \Big) + \sum_{l=1}^{L}\tau_{l} \tilde u_{i,t-l}^{2} + \sum_{k=1}^{K} \nu_{k} \tilde h_{i,t-k,\zeta, \tilde\omega_i},
\end{align*}
where
\begin{align*}
\check\omega_i = \frac{1}{T}\sum_{t=1}^{T}\check{u}_{it}^2,\,\,\,\,\,\check{u}_{it}=\tilde u_{it, \hat\lambda},\,\,\,\,\,
\tilde \omega_i = \frac{1}{T}\sum_{t=1}^{T} \tilde u_{it}^2,\,\,\,\mbox{ and }\,\,\,\tilde u_{it}=\tilde u_{it,\lambda_0}.
\end{align*}
Here, $\tilde{u}_{it,\lambda} = {y}_{it}-\tilde\mu_{i,\lambda} - {x}_{it}'\beta-\sum_{p=1}^{P}\phi_{p}{y}_{i,t-p}-\sum_{q=1}^{Q}\psi_{q}\tilde{u}_{i,t-q,\lambda}$
with $\tilde\mu_{i,\lambda} = (1 + \sum_{q=1}^{Q}\psi_{q}) T^{-1} $ $\sum_{t=1}^{T} (1 + \sum_{q=1}^{Q}\psi_{q}\mathcal{B}^q)^{-1}(y_{it} - x_{it}'\beta -\sum_{p=1}^{P}\phi_{p}{y}_{i,t-p})$ and $\mathcal{B}$ being the lag operator,
and $\tilde u_{it}$ can be re-written as  $\tilde u_{it}= u_{it} - T^{-1}\sum_{t=1}^{T}u_{it}$. Note that similar to $\hat\mu_{i,\lambda}$ in (\ref{equ_hat_mu}),
$\tilde\mu_{i,\lambda}$ is the solution of $\mu_{i}$ by solving the equations $\partial (U_{\theta_1}'U_{\theta_1}) /\partial \mu = 0$.
% with $Q_{\theta_1} = U_{\theta_1}'U_{\theta_1}$

Denote $\hat G := \partial \hat L_{\zeta, \hat\omega}/\partial \zeta\big|_{\zeta=\zeta_0}$.
First, by letting $\check\omega=(\check\omega_1,...,\check\omega_N)'\in\mathcal{R}^{N}$, we re-write
\begin{equation}\label{decom_1}
\hat G=\check{G}+G^{\dag},
\end{equation}
where $G^{\dag}:=\hat G-\check{G}$ captures the effect of unobservable initial values in ARMA and GARCH specifications, and $\check{G}$ is defined in the same way as $\hat G$ with
$\hat u_{it}$ and $\hat{h}_{it,\zeta_0,\hat{\omega}_i}$ replaced by $\check{u}_{it}$ and $\check{h}_{it,\zeta_0,\check{\omega}_i}$, respectively. Second, by letting $\tilde\omega=(\tilde \omega_1,...,\tilde \omega_N)'\in\mathcal{R}^{N}$, we re-write
\begin{equation}\label{decom_2}
\check{G}=\tilde{G}+G^{\S},
\end{equation}
where $G^{\S}:=\check{G}-\tilde{G}$ reflects the effect of the estimation of $\lambda_0$ by $\hat\lambda$, and $\tilde{G}$ is defined in the same way as $\hat G$ with
$\hat u_{it}$ and $\hat{h}_{it,\zeta_0,\hat{\omega}_i}$ replaced by $\tilde{u}_{it}$ and $\tilde{h}_{it,\zeta_0,\tilde\omega_i}$, respectively. Third, by letting $\bar\omega=(\bar \omega_1,...,\bar \omega_N)'\in\mathcal{R}^{N}$ with $\bar \omega_i =T^{-1}\sum_{t=1}^{T}  u_{it}^2$, we re-write
\begin{equation}\label{decom_3}
\tilde{G}=\bar{G}+G^{\sharp},
\end{equation}
where $G^{\sharp}:=\tilde{G}-\bar{G}$ considers the effect of the estimation of $\mu_0$ by $\tilde{\mu}_{\lambda_0}$, and $\bar{G}$ is defined in the same way as $\hat G$ with
$\hat u_{it}$ and $\hat{h}_{it,\zeta_0,\hat{\omega}_i}$ replaced by $u_{it}$ and $h_{it,\zeta_0,\bar \omega_i}$, respectively.
Finally, we re-write
\begin{equation}\label{decom_4}
\bar{G}=G+G^{\flat},
\end{equation}
where $G^{\flat}:=\bar{G}-G$ gives the effect of the estimation of $\omega_0$ by $\bar\omega$, and $G$ is defined in the same way as $\hat G$ with
$\hat u_{it}$ and $\hat{h}_{it,\zeta_0,\hat{\omega}_i}$ replaced by $u_{it}$ and $h_{it,\zeta_0,\omega_{0i}}$, respectively.

Now, denote
$$D - {\rm E}( D):=\sum_{i=1}^{N}\sum_{t=1}^{T}d_{it}\,\,\, \mbox{ and }\,\,\,\Psi_{it} : = \frac{1}{2}\Big[\frac{1}{h_{it}^2}\frac{\partial h_{it}}{\partial \zeta} - {\rm E}\Big( \frac{1}{h_{it}^2} \frac{\partial  h_{it}}{\partial \zeta}\Big)\Big](u_{it}^2 - h_{it}) + \frac{1}{2}{\rm E}(\mathcal{O}^{\sharp}_{23,it})u_{it},$$
where ${\rm E}(\mathcal{O}^{\sharp}_{23,it})$ is defined in the supplementary materials and it is zero if $\epsilon_{it}$ is systematic about zero. By (\ref{decom_1})--(\ref{decom_4}) and the proof in the supplementary materials, we have
\begin{align}
 \frac{1}{\sqrt{NT}}\hat G &= \frac{1}{\sqrt{NT}} G + \frac{1}{\sqrt{NT}} G^{\flat} +  \frac{1}{\sqrt{NT}} G^{\sharp} + \frac{1}{\sqrt{NT}} G^{\S} +\frac{1}{\sqrt{NT}} G^{\dag} \nonumber \\
&= \frac{1}{\sqrt{NT}} \sum_{i = 1}^{N}\sum_{t=1}^{T} \Big(\Psi_{it} +\Pi \Gamma_{1}^{-1}d_{it}\Big)
+ \frac{1}{\sqrt{NT}}\big(\Delta^{\flat} + \Delta^{\sharp} +   \Delta^{\S} +  \Delta^{\dagger}\big) + o_p(1) \nonumber\\
%&\qquad + O_p\Big( \max\Big(\sqrt{\frac{N}{T^2}}, \sqrt{\frac{1}{T}}\Big)\Big) \nonumber\\
&= O_p(1)  +  O\Big(\sqrt{\frac{N}{T}}\Big) + o_p(1). \label{equ_G_op}
\end{align}
%where
%\begin{equation*}
%\Psi_{it} = \Big[\frac{1}{h_{it}^2}\frac{\partial h_{it}}{\partial \zeta} +  {\rm E}\Big( \frac{1}{h_{it}^2} \frac{\partial  h_{it}}{\partial \zeta}\Big)\Big](u_{it}^2 - h_{it})
%\end{equation*}
In (\ref{equ_G_op}), the first item (resulting from $G$,  $G^{\flat}$, $G^{\sharp}$, and $G^{\S}$) gives the asymptotic normality of $\hat\zeta$, the second item (resulting from $G^{\flat}$, $G^{\sharp}$, $G^{\S}$, and $G^{\dagger}$) causes the asymptotic bias of $\hat\zeta$, and the last item is negligible.

%\Pi \Gamma_{1}^{-1}\big[{\rm E}(D) + {\rm E}(D^{\dagger})\big]

%For the reduced panel AR-ARCH model, the initial value problem would disappear only if the summation in $\hat L_{\zeta,\hat\omega}$ begins from $t= K+1$.

Denote $\hat W := \partial^2 \hat L_{\zeta, \hat\omega}/\partial \zeta\partial\zeta'\big|_{\zeta=\zeta_0}$. Similar to (\ref{equ_G_op}), from the proof in the supplementary materials, we can obtain
\begin{equation}\label{equ_W_op}
\frac{1}{NT}\hat W = \Gamma_{2} + O_p\Big(\sqrt{\frac{1}{T}}\Big) = O_p(1) +  o_p(1),
\end{equation}
where
$$\Gamma_{2} = \frac{1}{N}\sum_{i=1}^{N} {\rm E} \big( \Phi_{it}\Phi_{it}' \big) \mbox{ with }\Phi_{it} = \frac{1}{2h_{it}}\frac{\partial  h_{it}}{\partial \zeta}.$$
In (\ref{equ_W_op}), the first item contributes to the asymptotic variance of $\hat\zeta$, and the second item is negligible.

Define
$$\Xi = \frac{1}{N}\sum_{i=1}^{N}{\rm E}\big(\Psi_{it}\Psi_{it}'\big)\,\,\,\,\mbox{ and }\,\,\,\,
\Lambda = \frac{1}{N}\sum_{i=1}^{N}{\rm E}\big(d_{it}\Psi_{it}\big).$$
By (\ref{equ_G_op})--(\ref{equ_W_op}), we are ready to provide the asymptotics of $\hat\zeta$.

\begin{thm}\label{thm_hat_zeta}
Suppose that Assumptions \ref{ass_parameters}--\ref{ass_TN} hold. Then,
$
\hat\zeta\stackrel{\text{p}}{\longrightarrow}\zeta_0.
$
If Assumption \ref{ass_Gamma_Omega} also holds,
\begin{equation*} %\label{equ_asym_hat_zeta}
\sqrt{NT}\Big(\hat\zeta -  \zeta_0  - \frac{1}{T} c_{2} - \frac{1}{T} c_{2}^{\S} - \frac{1}{T} c_{2}^{\dagger} \Big)
\stackrel{\text{d}}{\longrightarrow} N(0,\Sigma_{2}),
\end{equation*}
where
\begin{align*}
& c_{2} =  \frac{\Gamma_{2}^{-1}(\Delta^{\flat} + \Delta^{\sharp})}{N} = O(1),\,\,\,\, c_{2}^{\S} = \frac{\Gamma_{2}^{-1}\Delta^{\S}}{N} = O(1),\,\,\,\,c_{2}^{\dag} = \frac{\Gamma_{2}^{-1}\Delta^{\dagger}}{N} = O(1),\\
&\mbox{and }\Sigma_{2} =  \lim \Gamma_{2}^{-1} \Omega_{2} \Gamma_{2}^{-1} \mbox{ with }\Omega_{2} =  \Xi +  \Pi\Gamma_{1}^{-1} \Omega_{1} \Gamma_{1}^{-1}\Pi' +  \Pi\Gamma_{1}^{-1}\Lambda  + \Lambda' (\Gamma_{1}^{-1})' \Pi'.
\end{align*}
\end{thm}

\begin{rem}\label{rem_zeta}
For saving the space, we give the explicit formulas of $\Delta^{\flat}$, $\Delta^{\sharp}$, $\Delta^{\S}$, $\Gamma_{2}$, $\Pi$, and $\Omega_{2}$ in Section B.2 of the supplementary materials. Then, under the conditions of Theorem \ref{thm_hat_zeta}, we can consistently estimate
$c_{2}$, $c_{2}^{\S}$, and $\Sigma_{2}$ by their plug-in counterparts $\hat c_{2}$, $\hat c_{2}^{\S}$, and $\hat\Sigma_{2}$, respectively.
For $\Delta^{\dagger}$, we show that $\Delta^{\dagger}=O(N)$ in Section C.3 of the supplementary materials, but we
are unable to provide its explicit formula due to the non-linearity of GARCH specification with some non-zero parameters $\nu_{0k}$. Consequently, we cannot offer a consistent estimator of $c_{2}^{\dag}$.
\end{rem}

%(\ref{equ_20231214})--(\ref{equ_hat_L_tau})

The results of Theorem \ref{thm_hat_zeta} are new to the literature, and they demonstrate that $\hat\zeta$ has a similar asymptotic behavior as $\hat\lambda$ in Theorem \ref{thm_lambda_consistent_asymptotic}. However, it is worth noting that the asymptotic bias of $\hat\zeta$ consists of three parts: $c_{2}/T$, $c_{2}^{\S}/T$, and $c_{2}^{\dagger}/T$. Specifically, the first part
$c_{2}/T$ comes from the estimated fixed effects $\hat\omega$ and $\hat\mu_{\hat\lambda}$,
the second part $c_{2}^{\S}/T$ stems from the bias of estimated ARMA parameter vector $\hat\lambda$, and the third part $c_{2}^{\dagger}/T$
results from the unobservable initial values in both panel ARMA and panel GARCH specifications.
Particularly, if $Q=0$ and $K=0$ (i.e., $y_{it}$ follows a panel AR--ARCH model), we can show that $\Delta^{\dagger}=0$, so the third part of the asymptotic bias of $\hat\zeta$ disappears.

Finally, we give a theorem to show that $\hat\omega_i$ and $\hat\varpi_i$ are $\sqrt{T}$-consistent and asymptotically normal without any asymptotic bias.

\begin{thm}\label{thm_omega_varpi}
Suppose that Assumptions \ref{ass_parameters}--\ref{ass_Gamma_Omega} hold. Then, for each fixed $i$,
\begin{equation*}
\sqrt{T} \big(\hat\omega_i - \omega_{0i}\big) \stackrel{\text{d}}{\longrightarrow} N(0, \sigma^2_{2i})\,\,\,\mbox{ and }
\,\,\,\sqrt{T} \big(\hat\varpi_i - \varpi_{0i}\big) \stackrel{\text{d}}{\longrightarrow} N(0, \bar{\sigma}_{2i}^2),
\end{equation*}
where
\begin{align*}
\sigma^2_{2i} &= \Big(\frac{1 - \sum_{k=1}^{K} \nu_{0k}}{1-\sum_{l=1}^{L}\tau_{0l} - \sum_{k=1}^{K} \nu_{0k}} \Big)^2 [{\rm E}(\epsilon_{it}^4) - 1] {\rm E}(h_{it}^2),\\
\bar{\sigma}_{2i}^2 &= \Big( 1 - \sum_{k=1}^{K} \nu_{0k}\Big)^2 [{\rm E}(\epsilon_{it}^4) - 1] {\rm E}(h_{it}^2).
\end{align*}
\end{thm}

\subsubsection{Bias correction of $\hat\zeta$}

Since the asymptotic bias $c_{2}^{\S}/T$ in Theorem \ref{thm_hat_zeta} is caused by that of $\hat\lambda$,
 we can exclude this asymptotic bias by constructing an alternative VT-QML estimator $\hat\zeta^{\star}$, which is formed in the same way as
 $\hat\zeta$ with $\hat\lambda$ replaced by its bias-corrected counterpart $\hat\lambda_A$ or $\hat\lambda_J$.
% Moreover, we can further remove the asymptotic bias $c_2/T$ in Theorem \ref{thm_hat_zeta} via the following
% analytic bias correction estimator
% \begin{equation}\label{equ_zeta_A}
%\hat\zeta_{A} = \hat\zeta^{\star}  - \frac{1}{T} \hat c_2,
%\end{equation}
%where $\hat c_2$ in Remark \ref{rem_zeta} is the plug-in estimator for $c_2$.
Unfortunately, the remaining asymptotic biases in Theorem \ref{thm_hat_zeta} cannot be eliminated by this analytical method, as their explicit formulas are unavailable; see the discussions in Remark \ref{rem_zeta} above. To deal with this problem, we follow Section \ref{subsection_asymptotics_mean} to consider the half-panel Jackknife bias correction estimator
\begin{equation}\label{equ_zeta_J}
\hat\zeta_{J} = 2\hat\zeta^{\star} - \frac{1}{2}(\hat\zeta^{\star}_1 + \hat\zeta^{\star}_2),
\end{equation}
where $\hat\zeta^{\star}_1$ and $\hat\zeta^{\star}_2$ are defined in the same way as $\hat\zeta^{\star}$, but only involving observations during $t\in \{1,2,..., \lfloor T/2 \rfloor\}$ and $t\in \{\lfloor T/2 \rfloor + 1,...,T\}$, respectively.
Now, we give the asymptotic normality of $\hat\zeta_{J}$ below:

\begin{thm}\label{thm_zeta_ddagger}
Suppose that Assumptions \ref{ass_parameters}--\ref{ass_Gamma_Omega} hold. Then,
\begin{align*}
%\sqrt{NT}\Big(\hat\zeta_{A} - \zeta_0 - \frac{1}{T} c_2^{\dagger} \Big)\stackrel{\text{d}}{\longrightarrow} N(0, \Sigma_2)
%\,\,\,\mbox{ and }\,\,\,
\sqrt{NT}(\hat\zeta_{J} - \zeta_0)\stackrel{\text{d}}{\longrightarrow} N(0, \Sigma_2).
\end{align*}
\end{thm}

The above theorem demonstrates that $\hat\zeta_{J}$ is $\sqrt{NT}$-consistent and asymptotically centered normal. Thus, it could have a better finite sample performance than $\hat\zeta$ in terms of bias, as shown by our simulation studies in the next section.

\section{Simulations}\label{section_simulation}
In this section, we examine the finite sample performance of the estimators $\hat\lambda$ and $\hat\zeta$ in (\ref{equ_hat_lambda}) and (\ref{equ_hat_zeta}), together with their analytical bias correction estimator  $\hat\lambda_A$ in (\ref{equ_lambda_A}), and Jackknife bias correction estimators $\hat\lambda_J$ and $\hat\zeta_J$ in (\ref{equ_lambda_J}) and (\ref{equ_zeta_J}).

Specifically, we generate 1000 replications of the sample size $N\times T$ from the following panel ARMA($1, 1$)--GARCH($1, 1$) model:
\begin{align}\label{equ_ARMAGARCH11}
\begin{split}
y_{it} &= \mu_{0i} + \beta_0 x_{it} + \phi_0 y_{i,t-1}  + \psi_0 u_{i,t-1}+u_{it},\\
u_{it} &=\sqrt{h_{it}}\epsilon_{it} \text{ with } h_{it} =  \omega_{0i}(1-\tau_0 - \nu_0) + \tau_0 u_{i,t-1}^2 +  \nu_0 h_{i,t-1},
\end{split}
\end{align}
where $N\in\{50, 100\}$, $T\in\{20, 50, 100, 200, 300\}$, $\lambda_0  = (\beta_0, \phi_0, \psi_0)' = (3,0.3,0.3)'$, $\zeta_0 = (\tau_0, \nu_0)' = (0.2,0.4)'$, $\mu_{0i}\overset{i.i.d.}{\sim} N(0, 1)$, $x_{it}\overset{i.i.d.}{\sim} N(0, 1)$, $\omega_{0i}\overset{i.i.d.}{\sim} U(1,3)$,
$\epsilon_{it}\overset{i.i.d.}{\sim} N(0, 1)$, and $\mu_{0i}$, $x_{it}$, $\omega_{0i}$, and $\epsilon_{it}$ are independent.
For each replication, we compute all of considered estimators
$\hat \lambda=(\hat\beta, \hat\phi, \hat\psi)'$, $\hat\zeta=(\hat\tau,\hat\nu)'$,  $\hat \lambda_{A}=(\hat\beta_A, \hat\phi_A, \hat\psi_A)'$,  $\hat \lambda_{J}=(\hat\beta_J, \hat\phi_J, \hat\psi_J)'$, and $\hat\zeta_J=(\hat\tau_J,\hat\nu_J)'$.
Based on the results from 1000 replications, Tables \ref{table_Bias_norm}, \ref{table_SD_norm}, and \ref{table_SDAD_norm} report the sample bias, the sample standard deviations (SD), and the ratio of SD over the average estimated
asymptotic standard deviations (AD) of all considered estimators, respectively.
Here, since the sample size $T = 20$ is too small to provide reliable estimation results for GARCH parameters,
the related results are excluded. From Tables \ref{table_Bias_norm}--\ref{table_SDAD_norm}, we can have the following findings:
\begin{itemize}
\item[(i)] For each given $N$, the biases of $\hat \lambda$ and $\hat\zeta$ decrease with the value of $T$. In contrast,  for each given $T$, the biases of  $\hat \lambda$ and $\hat\zeta$ do not have a clear decreasing trend when the value of $N$ increases. This observation matches our theoretical results in Theorems \ref{thm_lambda_consistent_asymptotic} and \ref{thm_hat_zeta} that the asymptotic bias of $\hat \lambda$ or $\hat\zeta$ has the order $O(T^{-1})$ which is irrespective of the value of $N$. Moreover, the bias-corrected estimators
$\hat \lambda_{A}$ and $\hat \lambda_{J}$ (or $\hat\zeta_J$) have much smaller biases than $\hat\lambda$ (or $\hat\zeta$), especially when the value of $T$ is small. This is consistent with our asymptotic analysis in Theorems \ref{thm_lambda_bias_cor} and \ref{thm_zeta_ddagger}.
In addition, compared with $\hat \lambda_{J}$, $\hat \lambda_{A}$ has slightly smaller biases when
$T$ is small, but this mild advantage disappears when $T$ is large.

% Hence, we recommend the analytical bias correction method for practical use.

\item[(ii)] The values of SD for all considered estimators become smaller when the value of either $N$ or $T$ increases, and they are nearly unchanged to the bias correction implementation. This supports our theoretical results in Theorems \ref{thm_lambda_consistent_asymptotic}, \ref{thm_lambda_bias_cor}--\ref{thm_hat_zeta}, and \ref{thm_zeta_ddagger} that all estimators are $\sqrt{NT}$-consistent and the formula of their asymptotic variance remains the same, irrespective of the bias correction implementation.

\item[(iii)] The values of SD/AD are close to one for all considered estimators, except the GARCH parameter estimators
$\hat\nu$  and $\hat\nu_{J}$ in the case of $T=50$. This finding is not unexpected, since the experience in the literature shows that it is more hard to estimate $\nu$ than $\tau$. It is worth noting that for classical GARCH models, a sample size as small as 100 cannot deliver accurate GARCH parameter estimators with reliable standard errors. Hence, although the estimators of $\nu$ are less satisfactory for $T=50$, our findings are still encouraging, since they indicate that the asymptotic normality of our considered estimators in Theorems \ref{thm_lambda_consistent_asymptotic} and \ref{thm_hat_zeta} holds even when the value of $T$ is around 100. Clearly, the panel structure of our model leads to the advantage of our estimators in dealing with moderate $T$ samples.
\end{itemize}

%$\sqrt{0.6}t(5)$

Overall, the above simulation results demonstrate that
$\hat\lambda_A$, $\hat\lambda_J$, and $\hat\zeta_J$ perform better than their competitors in terms of bias, and their asymptotic normality holds even when $T$ is as small as 100. In the supplementary materials, some additional simulation results for model (\ref{equ_ARMAGARCH11}) with student $t$ distributed errors are also provided, and they give us a similar conclusion as above.

\begin{table}[!ht]
\centering
\caption{Estimation results of bias for model (\ref{equ_ARMAGARCH11}) with $\epsilon_{it}\overset{i.i.d.}{\sim} N(0, 1)$.}
\footnotesize
\setlength{\tabcolsep}{0.5mm}  % 0.4
\renewcommand{\arraystretch}{1.9}   %0.86
\begin{tabular}{llccccccccccccc}		%rrrrrrrrrrrrrrr
\hline
\multicolumn{15}{c}{Bias}\\	
\cmidrule(r){3-15}
$N$&$T$&$\hat\beta$&$\hat\beta_A$&$\hat\beta_J$&$\hat\phi$&$\hat\phi_A$&$\hat\phi_J$&$\hat\psi$&$\hat\psi_A$&$\hat\psi_J$&$\hat\tau$&$\hat\tau_J$&$\hat\nu$&$\hat\nu_J$\\
\hline
50&20&$-$0.008&$-$0.001&0.003&$-$0.008&$-$0.001&0.005&$-$0.033&0.003&0.023&-----&-----&-----&-----\\
&50&$-$0.003&0.000&0.001&$-$0.004&$-$0.001&0.001&$-$0.012&0.000&0.005&$-$0.055&0.001&$-$0.135&0.024\\
&100&$-$0.001&0.001&0.001&$-$0.002&0.000&0.001&$-$0.005&0.001&0.003&$-$0.025&0.000&$-$0.053&0.022\\
&200&$-$0.001&0.000&0.000&$-$0.001&0.000&0.000&$-$0.003&0.000&0.001&$-$0.013&$-$0.001&$-$0.023&0.006\\
&300&$-$0.001&0.000&0.000&$-$0.001&0.000&0.000&$-$0.001&0.001&0.001&$-$0.008&0.000&$-$0.016&0.002\\
\hline
100&20&$-$0.008&$-$0.002&0.003&$-$0.008&$-$0.001&0.005&$-$0.032&0.004&0.023&-----&-----&-----&-----\\
&50&$-$0.004&$-$0.001&0.001&$-$0.004&0.000&0.001&$-$0.011&0.001&0.006&$-$0.054&0.003&$-$0.130&0.047\\
&100&$-$0.002&0.000&0.000&$-$0.002&0.000&0.001&$-$0.005&0.000&0.003&$-$0.026&0.000&$-$0.053&0.021\\
&200&0.000&0.001&0.001&$-$0.001&0.000&0.000&$-$0.003&0.000&0.001&$-$0.013&$-$0.001&$-$0.022&0.008\\
&300&0.000&0.001&0.001&$-$0.001&0.000&0.000&$-$0.002&0.000&0.001&$-$0.008&$-$0.001&$-$0.016&0.002\\
\hline						
\end{tabular}		
\label{table_Bias_norm}
\end{table}

\begin{table}[!ht]
\centering
\caption{Estimation results of SD for model (\ref{equ_ARMAGARCH11}) with $\epsilon_{it}\overset{i.i.d.}{\sim} N(0, 1)$.}
\footnotesize
\setlength{\tabcolsep}{1.4mm}  % 0.4
\renewcommand{\arraystretch}{1.9}   %0.86
\begin{tabular}{llccccccccccccc}		
\hline
\multicolumn{15}{c}{SD}\\	
\cmidrule(r){3-15}
$N$&$T$&$\hat\beta$&$\hat\beta_A$&$\hat\beta_J$&$\hat\phi$&$\hat\phi_A$&$\hat\phi_J$&$\hat\psi$&$\hat\psi_A$&$\hat\psi_J$&$\hat\tau$&$\hat\tau_J$&$\hat\nu$&$\hat\nu_J$\\
\hline
50&20&0.046&0.046&0.049&0.016&0.016&0.018&0.044&0.041&0.048&-----&-----&-----&-----\\
&50&0.028&0.028&0.028&0.010&0.010&0.010&0.027&0.027&0.029&0.028&0.035&0.101&0.157\\
&100&0.020&0.020&0.020&0.007&0.007&0.007&0.019&0.019&0.019&0.020&0.022&0.063&0.076\\
&200&0.015&0.014&0.015&0.005&0.005&0.005&0.013&0.012&0.013&0.014&0.015&0.044&0.048\\
&300&0.012&0.012&0.012&0.004&0.004&0.004&0.010&0.010&0.010&0.012&0.013&0.035&0.037\\
\hline
100&20&0.032&0.032&0.035&0.012&0.012&0.013&0.032&0.030&0.035&-----&-----&-----&-----\\
&50&0.020&0.020&0.021&0.007&0.007&0.007&0.019&0.018&0.020&0.020&0.024&0.073&0.114\\
&100&0.015&0.015&0.015&0.005&0.005&0.005&0.013&0.013&0.014&0.014&0.015&0.045&0.052\\
&200&0.010&0.010&0.010&0.004&0.004&0.004&0.009&0.009&0.009&0.010&0.011&0.031&0.033\\
&300&0.008&0.008&0.008&0.003&0.003&0.003&0.007&0.007&0.007&0.008&0.009&0.025&0.026\\
\hline						
\end{tabular}	
\label{table_SD_norm}										
\end{table}

\begin{table}[!ht]
\centering
\caption{Estimation results of SD/AD for model (\ref{equ_ARMAGARCH11}) with $\epsilon_{it}\overset{i.i.d.}{\sim} N(0, 1)$.}
\footnotesize
\setlength{\tabcolsep}{1.4mm}  % 0.4
\renewcommand{\arraystretch}{1.9}   %0.86
\begin{tabular}{llccccccccccccc}		
\hline
\multicolumn{15}{c}{SD/AD}\\	
\cmidrule(r){3-15}
$N$&$T$&$\hat\beta$&$\hat\beta_A$&$\hat\beta_J$&$\hat\phi$&$\hat\phi_A$&$\hat\phi_J$&$\hat\psi$&$\hat\psi_A$&$\hat\psi_J$&$\hat\tau$&$\hat\tau_J$&$\hat\nu$&$\hat\nu_J$\\
\hline
50&20&1.025&1.021&1.106&1.017&1.018&1.153&0.920&0.866&1.020&-----&-----&-----&-----\\
&50&0.998&0.997&1.010&1.001&1.001&1.043&1.022&1.003&1.083&1.022&1.279&0.737&1.976\\
&100&0.986&0.986&0.991&0.993&0.993&0.999&1.017&1.009&1.046&0.968&1.079&0.856&1.332\\
&200&1.034&1.034&1.045&0.955&0.955&0.981&0.967&0.963&0.988&0.973&1.048&0.950&1.163\\
&300&1.010&1.010&1.015&1.001&1.001&1.004&0.958&0.955&0.968&1.000&1.047&0.975&1.089\\
\hline
100&20&1.032&1.028&1.107&1.040&1.045&1.152&0.943&0.887&1.052&-----&-----&-----&-----\\
&50&1.020&1.019&1.047&1.011&1.012&1.058&0.981&0.962&1.029&1.030&1.239&0.773&2.128\\
&100&1.041&1.041&1.044&1.012&1.012&1.030&1.003&0.994&1.036&0.956&1.076&0.875&1.287\\
&200&1.022&1.022&1.021&1.047&1.047&1.044&0.959&0.954&0.973&1.004&1.066&0.949&1.142\\
&300&1.011&1.011&1.013&0.996&0.996&0.986&0.975&0.972&0.979&0.988&1.038&0.980&1.098\\
\hline						
\end{tabular}	
\label{table_SDAD_norm}										
\end{table}

\section{Empirical analysis}\label{section_example}

In this section, we use our panel ARMA--GARCH model to study the dynamics of Producer Price Indices (PPI) growth in domestic market for
31 different countries (see Table \ref{table_prediction} below for their names) over 120 months spanning from 2011.06 to 2021.06.
To investigate how other economic indicators affect the PPI, we also consider the Consumer Price Index (CPI) growth based on Energy or Food (CPI-E or CPI-F) and Industrial Production (IP) growth in Manufacturing or Construction (IP-M or IP-C) as four exogenous variables.
The data of all of these economic indicators are accessible from Organization for Economic Co-operation and Development (OECD)
at \href{https://www.oecd.org/}{https://www.oecd.org/}.

Let $y_{it}$, $x_{CPI\mbox{-}E, it}$, $x_{CPI\mbox{-}F, it}$, $x_{IP\mbox{-}M, it}$, and $x_{IP\mbox{-}C, it}$ denote the log difference of PPI, CPI-E,  CPI-F, IP-M, and IP-C growths of country $i$ at month $t$, respectively, where the last four exogenous variables are demeaned for a better interpretation on $y_{it}$. Based on these variables, we fit the PPI panel data $\{y_{it}\}$ via a panel ARMA($2, 2$)--GARCH($2, 2$) model in (\ref{equ_model_ARMA})--(\ref{equ_model_GARCH}), where $x_{it}=(x_{CPI\mbox{-}E, i,t-1}, x_{CPI\mbox{-}F, i,t-1}, x_{IP\mbox{-}M, i,t-1}, x_{IP\mbox{-}C, i,t-1})'$.
 After using the backward elimination procedure to remove those insignificant variables at the 5\% level, we obtain the following fitted
 ARMA($1, 1$)--GARCH($1, 1$) model:
 \begin{align}
	\label{empirical_model}
\begin{split}
y_{it} &= \hat\mu_i + 0.285y_{i,t-1} + 0.014x_{CPI\mbox{-}E, i,t-1} + 0.015 x_{IP\mbox{-}M, i,t-1} +  0.138 u_{i,t-1} + u_{it} ,\\
&\qquad\quad(0.057) \qquad\,\,\,\,(0.007)\qquad \qquad\quad\,\,\,(0.003) \qquad \qquad\,\,\,\,\,(0.059)\\
h_{it} &=   0.364 \hat\omega_i  +   0.223 u_{i,t-1}^2 + 0.413 h_{i,t-1},\\
&\qquad\qquad\quad\,\,\,(0.028) \qquad\,\,\,(0.074)
\end{split}
\end{align}
where the estimates are computed by the Jackknife bias-correction method, and the standard deviations of all estimates are given in parentheses.
 From this fitted model, we find that PPI has clear serial correlation and GARCH effect from the significance of the ARMA and GARCH parameters, respectively; moreover, we find that two variables CPI-E and IP-M have considerable positive influences on PPI, whereas the eliminated variables CPI-F and IP-C have no significant impact on PPI.

\begin{figure}[!ht]
	\centering
	\includegraphics[width=1\textwidth]{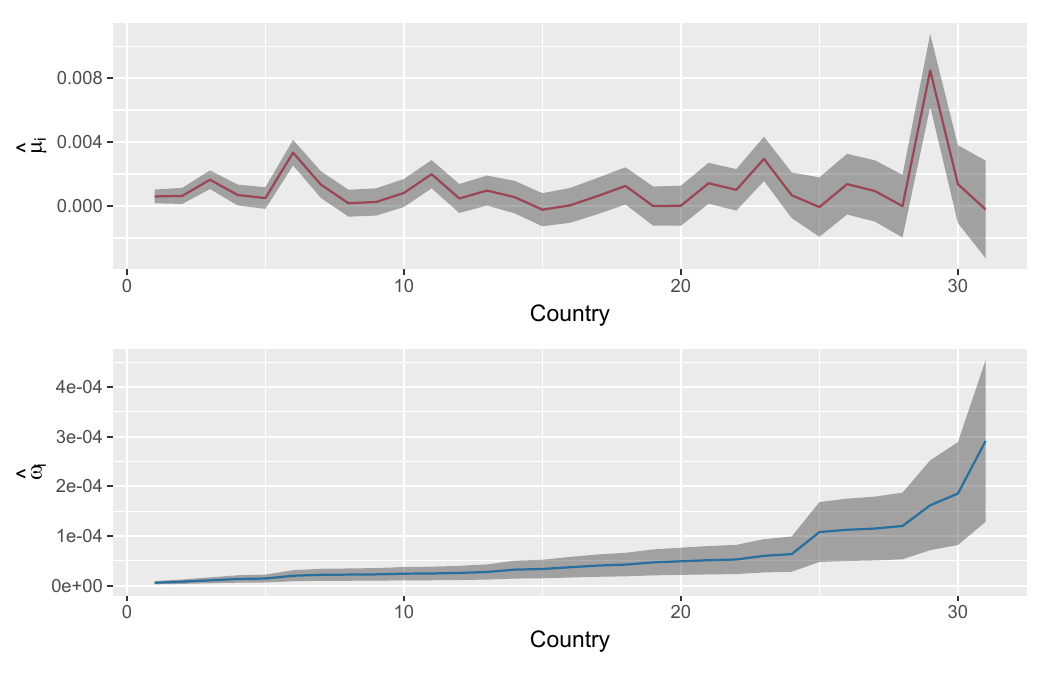}
	\caption{The estimated individual effects $\hat\mu_i$ and $\hat\omega_i$ with $95\%$ confidence intervals for 31 countries.}
	\label{figure_mu_omega}
\end{figure}

 Next, we examine the individual heterogeneity in both mean and variance by plotting the estimated individual effects $\hat\mu_i$ and $\hat\omega_i$ with $95\%$ confidence intervals in Figure \ref{figure_mu_omega}. Note that for each country $i$, the value of $\hat\mu_i$ (divided by $(1-0.285)$) represents the mean of $y_{it}$ due to the zero mean of $x_{CPI\mbox{-}E, i,t-1}$ and $x_{IP\mbox{-}M, i,t-1}$, while that of $\hat\omega_i$ represents the variance of  $y_{it}$. For ease of interpretation, we align 31 countries in a way such that
 the values of $\hat\omega_i$ are in ascending order, as shown in Figure \ref{figure_mu_omega}. From this figure, we observe
 the clear individual heterogeneity in both mean and variance.
 Specifically, we find that (i) except Countries `15', `19', `25', `28', and `31'  (Korea, Portugal, Israel, Lithuania, and Greece), all the values of $\hat\mu_i$ are positive, although some of them are not significantly different from zero; (ii)
 Countries `6',  `23', and `29' (South Africa,  Mexico, and Turkiye) have spiked values of $\hat\mu_i$, giving a sign of severe
 inflation in these three countries; and (iii) Countries `29',  `30', and `31' (Turkiye,  Hungary, and Greece) have much larger values of
 $\hat\omega_i$ than other countries, indicating that these three countries could have a higher risk of undergoing hyperinflation or hyperdeflation.

Finally, we assess the out-of-sample forecast performance of model (\ref{empirical_model}).
For this purpose, we use the following rolling window procedure: first, do the estimation based on the data of the latest 96 months to fit model (\ref{empirical_model}) and apply the fitted model to forecast the future $y_{it}$ as well as its two-sided 95\% confidence interval one month ahead; then repeat the foregoing procedure until the end of the sample has been reached. Based on the fitted model, the point forecast of
$y_{it}$ is computed in a conventional way, and the 95\% interval forecast of $y_{it}$ is computed by using the
filtered historical simulation method (\citealp{BaroneAdesi1998}). To see how the bias correction method affects the forecast performance, we compute the fitted model (\ref{empirical_model}) by using $(\hat{\lambda}', \hat{\zeta}')'$, $(\hat{\lambda}_{A}', \hat{\zeta}_{J}')'$, and
$(\hat{\lambda}_{J}', \hat{\zeta}_{J}')'$, leading to the so-called Methods I, I$_A$, and I$_J$, respectively. As a comparison, we
also consider Method II, which fits each $y_{it}$ based on the univariate ARMA($1, 1$)--GARCH($1, 1$) model
\begin{align}\label{empirical_model_one}
\begin{split}
y_{it} &= \mu_{0i} + \phi_{0i} y_{i,t-1} + \beta_{1,0i}x_{CPI-E,i,t-1}+  \beta_{2,0i}x_{IP-M,i,t-1}  + \psi_{0i} u_{i,t-1} + u_{it}, \\
u_{it} &=\sqrt{h_{it}}\epsilon_{it} \text{ with } h_{it} =  \omega_{0i}(1-  \tau_{0i} - \nu_{0i}) + \tau_{0i} u_{i,t-1}^2 +  \nu_{0i} h_{i,t-1},
\end{split}
\end{align}
where all of unknown parameters above are estimated equation by equation.

\begin{table}[p]
\centering
\caption{The RMSE for point forecasts and  p-value of $LR_{cc}$ for interval forecasts, based on Methods I, I$_A$, I$_J$, and II.}
\setlength{\tabcolsep}{2.5mm}
\renewcommand{\arraystretch}{1.2}
\begin{tabular}{cccccccccc}
\toprule
&&\multicolumn{4}{c}{RMSE}&\multicolumn{4}{c}{p-value of $LR_{cc}$}\\
\cmidrule(r){3-6}\cmidrule(r){7-10}
$i$&Country&I&I$_A$&I$_J$&II&I&I$_A$&I$_J$&II\\
\cmidrule(r){1-10}
1&Germany&0.347&0.325&\cellcolor{gray!30}0.320&0.341&\cellcolor{gray!30}0.629&\cellcolor{gray!30}0.629&\cellcolor{gray!30}0.629&0.003\\
2&Slovenia&0.469&0.470&\cellcolor{gray!30}0.461&0.486&0.001&\cellcolor{gray!30}0.005&\cellcolor{gray!30}0.005&0.000\\
3&Costa Rica&0.403&0.368&\cellcolor{gray!30}0.359&0.374&\cellcolor{gray!30}1.000&\cellcolor{gray!30}1.000&\cellcolor{gray!30}1.000&\cellcolor{gray!30}1.000\\
4&UK&0.352&\cellcolor{gray!30}0.345&0.346&\cellcolor{gray!30}0.345&\cellcolor{gray!30}0.945&\cellcolor{gray!30}0.945&\cellcolor{gray!30}0.945&\cellcolor{gray!30}0.945\\
5&Italy&0.485&0.449&\cellcolor{gray!30}0.442&0.593&\cellcolor{gray!30}0.202&0.190&0.190&0.012\\
6&South Africa&0.396&\cellcolor{gray!30}0.393&0.396&0.500&\cellcolor{gray!30}0.945&\cellcolor{gray!30}0.945&\cellcolor{gray!30}0.945&0.629\\
7&Latvia&0.702&0.657&\cellcolor{gray!30}0.642&0.822&0.016&\cellcolor{gray!30}0.202&\cellcolor{gray!30}0.202&0.000\\
8&Japan&0.570&\cellcolor{gray!30}0.567&\cellcolor{gray!30}0.567&0.578&0.210&\cellcolor{gray!30}0.629&\cellcolor{gray!30}0.629&\cellcolor{gray!30}0.629\\
9&France&0.586&0.576&\cellcolor{gray!30}0.575&0.586&0.629&\cellcolor{gray!30}0.945&\cellcolor{gray!30}0.945&\cellcolor{gray!30}0.945\\
10&Denmark&0.501&\cellcolor{gray!30}0.497&0.498&0.541&0.030&\cellcolor{gray!30}0.047&\cellcolor{gray!30}0.047&0.030\\
11&Colombia&0.509&0.502&\cellcolor{gray!30}0.501&0.507&\cellcolor{gray!30}0.945&\cellcolor{gray!30}0.945&\cellcolor{gray!30}0.945&\cellcolor{gray!30}0.945\\
12&Czechia&0.584&0.591&0.587&\cellcolor{gray!30}0.540&\cellcolor{gray!30}0.629&\cellcolor{gray!30}0.629&\cellcolor{gray!30}0.629&0.042\\
13&Poland&0.679&0.646&\cellcolor{gray!30}0.633&0.730&\cellcolor{gray!30}0.629&\cellcolor{gray!30}0.629&\cellcolor{gray!30}0.629&0.000\\
14&Finland&0.969&0.958&\cellcolor{gray!30}0.956&0.991&\cellcolor{gray!30}0.021&\cellcolor{gray!30}0.021&\cellcolor{gray!30}0.021&0.000\\
15&Korea&0.716&0.682&0.664&\cellcolor{gray!30}0.532&\cellcolor{gray!30}0.210&\cellcolor{gray!30}0.210&\cellcolor{gray!30}0.210&\cellcolor{gray!30}0.210\\
16&Ireland&\cellcolor{gray!30}0.410&0.445&0.448&0.452&0.307&0.307&0.307&\cellcolor{gray!30}0.945\\
17&Spain&0.913&0.873&0.868&\cellcolor{gray!30}0.857&\cellcolor{gray!30}0.629&\cellcolor{gray!30}0.629&\cellcolor{gray!30}0.629&0.202\\
18&Estonia&0.704&0.680&\cellcolor{gray!30}0.665&0.929&\cellcolor{gray!30}0.210&\cellcolor{gray!30}0.210&\cellcolor{gray!30}0.210&\cellcolor{gray!30}0.210\\
19&Portugal&0.737&0.712&0.711&\cellcolor{gray!30}0.686&0.042&\cellcolor{gray!30}0.210&\cellcolor{gray!30}0.210&\cellcolor{gray!30}0.210\\
20&Slovak Republic&0.755&0.713&\cellcolor{gray!30}0.709&0.857&\cellcolor{gray!30}0.629&\cellcolor{gray!30}0.629&\cellcolor{gray!30}0.629&0.210\\
21&Norway&\cellcolor{gray!30}0.813&0.817&0.829&0.888&\cellcolor{gray!30}0.629&\cellcolor{gray!30}0.629&\cellcolor{gray!30}0.629&0.210\\
22&Sweden&0.809&0.792&\cellcolor{gray!30}0.796&0.828&\cellcolor{gray!30}0.629&0.210&0.210&0.210\\
23&Mexico&1.026&1.016&1.023&\cellcolor{gray!30}0.996&\cellcolor{gray!30}0.945&\cellcolor{gray!30}0.945&\cellcolor{gray!30}0.945&\cellcolor{gray!30}0.945\\
24&Netherlands&0.925&0.885&\cellcolor{gray!30}0.875&0.964&0.210&0.210&0.210&\cellcolor{gray!30}0.629\\
25&Israel&1.286&\cellcolor{gray!30}1.269&1.271&1.296&\cellcolor{gray!30}0.629&\cellcolor{gray!30}0.629&\cellcolor{gray!30}0.629&\cellcolor{gray!30}0.629\\
26&Luxembourg&1.096&\cellcolor{gray!30}1.068&1.072&1.129&\cellcolor{gray!30}0.047&\cellcolor{gray!30}0.047&\cellcolor{gray!30}0.047&\cellcolor{gray!30}0.047\\
27&Belgium&1.486&1.437&\cellcolor{gray!30}1.432&2.234&\cellcolor{gray!30}0.629&\cellcolor{gray!30}0.629&\cellcolor{gray!30}0.629&0.202\\
28&Lithuania&1.586&1.537&\cellcolor{gray!30}1.519&1.589&\cellcolor{gray!30}0.089&\cellcolor{gray!30}0.089&\cellcolor{gray!30}0.089&0.021\\
29&Turkiye&1.317&1.234&1.188&\cellcolor{gray!30}1.106&\cellcolor{gray!30}1.000&\cellcolor{gray!30}1.000&\cellcolor{gray!30}1.000&\cellcolor{gray!30}1.000\\
30&Hungary&\cellcolor{gray!30}2.493&2.508&2.533&2.554&0.089&0.089&0.089&\cellcolor{gray!30}0.945\\
31&Greece&2.079&2.062&\cellcolor{gray!30}2.055&2.081&\cellcolor{gray!30}0.945&\cellcolor{gray!30}0.945&\cellcolor{gray!30}0.945&\cellcolor{gray!30}0.945\\
%\hline
%\multicolumn{2}{c}{Mean}&0.861&0.841&\cellcolor{gray!30}0.837&0.900&0.474&\cellcolor{gray!30}0.496 &\cellcolor{gray!30}0.496&0.418\\
\bottomrule		
\end{tabular}
\begin{tablenotes}  \footnotesize
\item \textit{Note}: For each country, the smallest RMSE and the largest p-value of $LR_{cc}$ (including ties) are highlighted in shadow.
\end{tablenotes} 				
\label{table_prediction}															
\end{table}

For each country $i$, Table \ref{table_prediction} reports its values of root mean squared error (RMSE) for point forecasts and its p-values of
conditional coverage test $LR_{cc}$ (\citealp{Christoffersen1998}) for 95\% interval forecasts, across four different methods.
From this table, we find that in terms of the minimized RMSE for point forecasts, (i) Method I$_J$ performs best, since it has a smaller RMSE
than Methods I, I$_A$, and II in 26, 21, and 26 countries, respectively; (ii) Method I$_A$ has the second best performance, with a
smaller RMSE than Methods I and II in 26 and 24 countries, respectively; (iii) although Method I performs worse than Methods I$_J$ and I$_A$, it still outperforms Method II with a smaller RMSE in 20 countries; (iv) Method II performs worst, as it has the smallest RMSE in only 7 countries (including ties). Moreover, in terms of the maximized p-value of $LR_{cc}$ for interval forecasts, we find that Methods I$_A$ and I$_J$ have the same performance, and they slightly outperform Method I with a larger p-value of $LR_{cc}$ in 6 countries; meanwhile, all of Methods I, I$_A$, and I$_J$ perform much better than Method II, which delivers invalid interval forecasts in 9 countries with the p-value of $LR_{cc}$ less than 5\%.

Overall, our findings from Table \ref{table_prediction} imply that model (\ref{empirical_model}) produces much better
point and interval forecasts than model (\ref{empirical_model_one}), and the former should be used with
the bias correction to improve the forecast performance. The advantage of model (\ref{empirical_model}) over model (\ref{empirical_model_one}) is probably caused by the fact that the estimators $(\hat{\lambda}', \hat{\zeta}')'$, $(\hat{\lambda}_{A}', \hat{\zeta}_{J}')'$, and
$(\hat{\lambda}_{J}', \hat{\zeta}_{J}')'$ for model (\ref{empirical_model}) are $\sqrt{NT}$-consistent due to its panel structure, whereas the estimators for model (\ref{empirical_model_one}) are only  $\sqrt{T}$-consistent and they thus are less reliable for smaller value of $T$.

\section{Conclusion}\label{section_conclusion}
In this paper, we propose a new panel ARMA--GARCH model.
This model captures the serial correlation in mean via a panel ARMA specification, the conditional heteroskedasticity in variance via a
panel GARCH specification, and the individual heterogeneity in both mean and variance via the unobservable fixed effects.
Using the concentration and VT treatments to solve the incidental parameter problem, we
construct the LS estimator for the panel ARMA specification and the VT-QML estimator for the panel GARCH specification stepwisely.
When both $N$ and $T$ diverge to infinity at the same rate,
we establish the asymptotic normality of both LS and VT-QML estimators, which have asymptotic biases of order $O(1/T)$ if
$N/T\not\to 0$. Specifically, we find that the presence of fixed effects and unobservable initial values leads to asymptotic biases of
both LS and VT-QML estimators, and the VT-QML estimator even has the asymptotic bias caused by the estimation of
panel ARMA specification. To correct the biases, we further propose the bias-corrected LS and VT-QML estimators by using either the analytical asymptotics or jackknife method. Simulations and one real example are given to illustrate the importance of the panel ARMA--GARCH model.
As an important theoretical development, we provide a new CLT for the linear-quadratic form in the martingale difference sequence, when the weight matrix is uniformly bounded in row and column. This CLT facilitates our technical proofs for the proposed LS and VT-QML estimators, and it could be useful for studying the estimation of complex models with time effects, spatial effects, or other cross-section effects in the future.

%%%%%%%%%%%%%%%%%%%%%%%%%%%%%%%%%%%%%%
%%%%%%%%%%%%%%%%%%%%%%%%%%%%%%%%%%%%%%

\end{document}